\documentclass[twocolumn,english,showpacs,superscriptaddress,preprintnumbers,amsmath,amssymb,floatfix]{revtex4-1}
\usepackage[T1]{fontenc}
\usepackage[latin9]{inputenc}
\usepackage{array}
\usepackage{amsmath}
\usepackage{graphicx}

\makeatletter

\providecommand{\tabularnewline}{\\}

\usepackage{latexsym}
\usepackage{dcolumn}
\usepackage{bm}
\usepackage{color}
\usepackage{babel}
\usepackage{amsfonts}
\usepackage{enumerate}
\usepackage{mathbbol}
\usepackage{braket}
\global\long\def\sgn{\mathrm{sgn}}

\makeatother

\usepackage{babel}
\begin{document}

\title{Parafermionic generalization of the topological Kondo effect}

\author{Kyrylo Snizhko}

\affiliation{Department of Condensed Matter Physics, Weizmann Institute of Science,
Rehovot, 76100 Israel}

\author{Francesco Buccheri}

\affiliation{Institut für Theoretische Physik, Heinrich-Heine-Universität, D-40225
Düsseldorf, Germany}

\author{Reinhold Egger}

\affiliation{Institut für Theoretische Physik, Heinrich-Heine-Universität, D-40225
Düsseldorf, Germany}

\author{Yuval Gefen}

\affiliation{Department of Condensed Matter Physics, Weizmann Institute of Science,
Rehovot, 76100 Israel}

\date{\today}
\begin{abstract}
We propose and study a parafermionic generalization of the topological
Kondo effect. The latter has been predicted to arise for a Coulomb-blockaded
mesoscopic topological superconductor (Majorana box), where at least
three normal leads are tunnel-coupled to different Majorana zero modes
on the box. The Majorana states represent a quantum impurity spin
that is partially screened due to cotunneling processes between leads,
with a stable non-Fermi liquid ground state. Our theory studies a
generalization where (i) Majorana states are replaced by topologically
protected parafermionic zero modes, (ii) charging effects again define
a spin-like quantum impurity on the resulting parafermion box, and
(iii) normal leads are substituted by fractional edge states. In this
multi-terminal problem, different fractional edge leads couple only
via the parafermion box. We show that although the linear conductance
tensor exhibits similar behavior as in the Majorana case, both at
weak and strong coupling, our parafermionic generalization is actually
not a Kondo problem but defines a rich new class of quantum impurity
problems. At the strong-coupling fixed point, a current injected through
a reference lead will be isotropically partitioned into outgoing currents
in all other leads, together with a universal negative current scattered
into the reference lead. The device can thus be operated as current
extractor, where the current partitioning is noiseless at the fixed
point. We describe a fractional quantum Hall setup proximitized by
superconductors and ferromagnets, which could allow for an experimental
realization in the near future. 
\end{abstract}
\maketitle

\section{Introduction}

A major goal of modern condensed matter physics is to understand and
predict the physics of Majorana zero modes \cite{Alicea2012,Leijnse2012,Beenakker2013,Lutchyn2018}
and their generalizations such as parafermionic zero modes \cite{Nayak2008,Alicea2016}.
Apart from the fundamental interest in observing, manipulating, and
controlling exotic fractionalized excitations, quantum states encoded
by sets of zero modes with non-Abelian braiding statistics hold significant
promise for quantum information processing applications due to their
topologically protected and highly nonlocal character \cite{Nayak2008,Sarma2015}.
While experimental evidence for Majorana states is rapidly mounting
in different platforms \cite{Mourik2012,Deng2012,Das2012,Rokhinson2012,Yazdani2014,Franke2015,Sun2016,Albrecht2016,Deng2016,Guel2017,Zhang2017,Albrecht2017,Nichele2017,Suominen2017,Gazi2017,Feldman2017,Deacon2017},
experimental searches for condensed-matter realizations of parafermions
(PFs) with symmetry $\mathbb{Z}_{n>2}$ are just about to start \cite{Ronen2018,Wu2018}.
(Note that $\mathbb{Z}_{2}$ PFs reduce to Majorana zero modes.)

The theoretical understanding of PFs, on the other hand, is already
comparatively well advanced, and many interesting phenomena have been
predicted \cite{Alicea2016,Lindner2012,Cheng2012,Clarke2013,Burrello2013,Vaezi2013,Zhang2014,Mong2014,Clarke2014,Barkeshli2014a,Barkeshli2014b,Klinovaja2014,Klinovaja2014b,Cheng2015,Alicea2015a,Kim2017,Snizhko2018,Pachos2018,Chew2018}.
While it has recently been shown that $\mathbb{Z}_{4}$ PFs admit
a free-fermion description \cite{Pachos2018,Chew2018}, theoretical
constructions for more general cases usually exploit the competition
between different gapping mechanisms at edge states of a topologically
ordered two-dimensional phase. Suggested platforms for hosting PF
zero modes include bilayer fractional quantum Hall (FQH) systems \cite{Barkeshli2014b},
proximitized fractional topological insulators \cite{Lindner2012},
and proximitized FQH liquids at filling factor $\nu=2/3$ \cite{Mong2014}
or $\nu=1/(2k+1)$ with integer $k$ \cite{Lindner2012,Clarke2013}.
Such setups have, in principle, the potential to ultimately realize
Fibonacci anyons capable of topologically protected universal quantum
computations \cite{Mong2014,Alicea2015a}. In particular, opposite-spin
FQH edges proximitized by alternating domains of superconductors (SCs)
and ferromagnets (FMs) should trap stable PF zero modes at domain
walls \cite{Lindner2012}. We note that recent experimental progress
has demonstrated that the seemingly conflicting requirements of high
magnetic fields (for the FQH phase) and superconductivity in principle
can be reconciled \cite{Lee2017,Lee2017a}.

In the present work, we analyze a previously unnoticed aspect of PFs
arising in the presence of Coulomb charging effects. In fact, recent
theoretical \cite{Fu2010,Zazunov2011,Hutzen2012} and experimental
\cite{Albrecht2016} work has highlighted the importance of Coulomb
charging effects in a floating mesoscopic superconductor hosting Majorana
bound states (`Majorana box'). By gapping out charge degrees of freedom
and by blocking detrimental processes related to quasiparticle poisoning
\cite{Aasen2016,Plugge2016}, a large box charging energy $E_{C}$
can further stabilize the Majorana subsector of the Hilbert space.
Importantly, charging effects will also activate long-range cotunneling
processes between different leads (or other access elements) attached
to the box via tunnel contacts. Consequently, Majorana boxes are key
ingredients for recently proposed topological quantum information
processing schemes \cite{Plugge2016,Landau2016,Plugge2017,Karzig2017,Litinski2017}.
When the Majorana box is operated under Coulomb valley conditions,
with $M\ge3$ normal-conducting (effectively spinless) leads tunnel-coupled
to Majorana zero modes on the box, the Majorana sector becomes equivalent
to an effective quantum impurity spin with SO$(M)$ symmetry \cite{Beri2012}.
For the minimal case with $M=3$, this impurity spin corresponds to
a standard spin-$1/2$ operator $(\hat{s}_{x},\hat{s}_{y},\hat{s}_{z})$,
where distinct operator components are nonlocally represented by different
Majorana bilinears on the box. Noting that also the leads have SO$(M)$
symmetry, cotunneling processes between different leads then act like
an exchange coupling and thus partially screen the effective impurity
spin. Ultimately, such processes drive the system to a robust non-Fermi
liquid fixed point analogous to the overscreened multi-channel Kondo
fixed point. For a detailed discussion of this topological Kondo effect
(TKE), see Refs.~\cite{Beri2012,Altland2013,Beri2013,Crampe2013,Tsvelik2013,Zazunov2014,Altland2014,Galpin2014,Buccheri2015,Zazunov2017}.
More generally, when the Majorana box is contacted by $M\ge3$ spinless
Luttinger liquid leads with interaction parameter $g$, where $g=1$
for the noninteracting case and $g<1$ ($g>1$) for repulsive (attractive)
electron-electron interactions \cite{Gogolin1998,Altland2010}, the
linear conductance between leads $j$ and $k$ is given by \cite{Beri2013,Zazunov2014}
\begin{equation}
G_{jk}^{{\rm TKE}}=\frac{2ge^{2}}{h}\left[1-\left(T/T_{K}\right)^{2\Delta_{M}-2}+\cdots\right]\left(\frac{1}{M}-\delta_{jk}\right),\label{tkecond}
\end{equation}
with the scaling dimension $\Delta_{M}=2g(M-1)/M$ of the leading
irrelevant operator at the TKE strong-coupling fixed point. This result
holds for $\Delta_{M}>1$ and temperatures $T$ well below the Kondo
temperature $T_{K}$. Apart from the non-Fermi-liquid power-law $T$
dependence, it is remarkable that the conductance tensor \eqref{tkecond}
is completely isotropic. Conductance measurements could thereby provide
strong evidence for nonlocality. For instance, putting $g=1$ and
$M=3$ in Eq.~\eqref{tkecond}, the $T=0$ conductance between leads
1 and 2 has the large value $G_{12}^{{\rm TKE}}=2e^{2}/3h$. If one
now decouples lead 3 from the box, e.g., by changing a gate voltage
to switch off the respective tunnel coupling, the TKE will be destroyed.
As a consequence, only an exponentially small conductance $G_{12}$
due to residual cotunneling is expected \cite{Hutzen2012}, without
the huge Kondo enhancement factor. This behavior is a clear signature
of nonlocality since the Majorana state coupled to lead 3 is centered
far away from the Majorana states coupled to leads 1 and 2.

In Ref.~\cite{Snizhko2018}, we have introduced a PF box device generalizing
the Majorana box to a setup with parafermionic zero modes. The PF
box of Ref.~\cite{Snizhko2018} could be realized in terms of opposite-spin
FQH edge states proximitized by alternating SC and FM domains, closely
following earlier proposals \cite{Lindner2012,Clarke2013} but taking
into account the box charging energy $E_{C}$. We emphasize that recent
experimental works have made significant steps towards implementing
such setups \cite{Ronen2018,Wu2018,Lee2017,Lee2017a}, and we are
confident that the model studied below can be realized in the near
future. The setup described in Ref.~\cite{Snizhko2018} also included
other access elements, in particular additional fractional edge states,
for readout and/or manipulation of the PF box state. The present work
is dedicated to studying a PF generalization of the Majorana-based
topological Kondo model, where the normal-conducting (Luttinger liquid)
leads behind Eq.~\eqref{tkecond} are replaced by FQH edge states,
see Fig.~\ref{fig1} for a schematic illustration of our setup. Such
leads correspond to chiral Luttinger liquids hosting fractional quasiparticles
\cite{Wen1991,Kane1992a,Gogolin1998,Altland2010}, and they have been
used experimentally for more than two decades \cite{Goldman1995,Goldman1996,Picciotto1997,Saminadayar1997,Goldman1997,Maasilta1997}.

In order to see whether PFs can establish a Kondo effect, we study
whether (and if yes, how) the TKE conductance tensor in Eq.~\eqref{tkecond}
is modified for a generalized PF setting. In fact, we find that the
PF generalization cannot in general be written as a Kondo Hamiltonian,
invariant under the action of a continuous group. For $M=3$ edges,
for example, we find that the 'quantum impurity spin' of the PF box
transforms in a representation of the SU$(n)$ group, where $n=2/\nu$
at filling factor $\nu=1/(2k+1)$. However, the low-temperature effective
Hamiltonian does not have this symmetry. This is related to the fact
that one cannot perform rotations in lead space since each FQH lead
necessitates different Klein factors. Nevertheless, we find that,
for $M\ge3$ chiral edge leads, a non-trivial strong-coupling regime
will be approached, where the conductance tensor exhibits an almost
identical behavior as for the TKE in Eq.~\eqref{tkecond}.

The PF generalization of the TKE thus constitutes a new type of multi-terminal
quantum junction distinct from previously studied cases \cite{Nayak1999,Chen2002,Chamon2003,Barnabe2005,Oshikawa2006,Das2006,Hou2008,Agarwal2009,Giuliano2009,Bellazzini2009,Altland2012,Rahmani2012,Altland2015}.
However, we remark that transport in the PF box case exhibits qualitative
(and technical) similarities to the TKE \cite{Altland2013,Beri2013,Zazunov2014}
as well as to the setup in Refs.~\cite{Altland2012,Altland2015}.
Although the physical realization and the detailed transport characteristics
differ, all three problems share several key features. In particular,
(i) the system is driven to a strong coupling regime, where (ii) an
incoming current is distributed between all the outgoing channels
in a nonlocal universal manner, and where (iii) this current partitioning
does not produce shot noise at the fixed point.

The remainder of this article is structured as follows. In Sec.~\ref{sec2},
we introduce our model for a PF generalization of the TKE. Abelian
bosonization allows one to solve the weak-coupling regime, where we
discuss the one-loop renormalization group (RG) equations in Sec.~\ref{sec3}.
Next, in Sec.~\ref{sec4}, we demonstrate that also the stable strong-coupling
fixed point can be accessed by Abelian bosonization. Related results
for the TKE have been obtained \cite{Altland2013,Beri2013} by using
an analogy to quantum Brownian motion in a periodic lattice \cite{Yi1998,Yi2002}.
We here instead employ the method of Ref.~\cite{Ganeshan2016}. Finally,
in Sec.~\ref{sec5}, we offer some conclusions. Technical details
have been delegated to several appendices, and we often put $\hbar=e=k_{B}=1$.

\section{Model}

\label{sec2}

\begin{figure}[t]
\centering \includegraphics[width=0.4\textwidth]{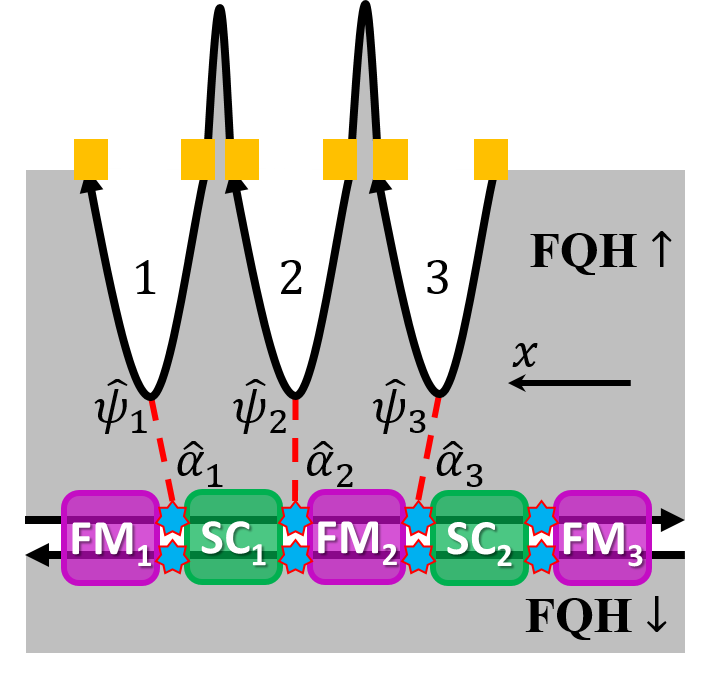} \caption{Schematic setup for a PF generalization of the topological Kondo effect.
Two opposite-spin FQH edges (thick straight black arrows) are gapped
out in different regions by the proximitizing FM and SC parts, where
PF zero mode operators $\hat{\alpha}_{j}$ (blue stars) are localized
at domain walls. (Strictly speaking, we have PF operators $\hat{\alpha}_{j}=\hat{\alpha}_{j,\uparrow}$
and $\hat{\alpha}_{j,\downarrow}$ at opposite-spin edges \cite{Snizhko2018}.
Here only the $\hat{\alpha}_{j}$ are needed.) The $N=2$ SC domains
are electrically connected to form a phase-coherent device with a
common charging energy $E_{C}$. Similarly, FM domains belong to one
bulk FM. The $M=3$ additional FQH edges (curved black arrows) serve
as quasiparticle leads, where the quasiparticle operator $\hat{\psi}_{j}$
is tunnel-coupled (red dashed lines) to the respective PF operator
$\hat{\alpha}_{j}$. Although different leads are parts of a single
long edge, they must be dynamically independent. To that end, Ohmic
contacts (yellow rectangles) are inserted between them. }
\label{fig1} 
\end{figure}

We start by discussing the Hamiltonian for our PF generalization of
the TKE. To keep the paper self-contained, we also include a brief
summary of those results of Ref.~\cite{Snizhko2018} needed below.
Following Ref.~\cite{Lindner2012}, we consider an array of PF zero
modes implemented via two $\nu=1/(2k+1)$ FQH puddles with opposite
spin polarization, see Fig.~\ref{fig1}. Related setups have recently
been achieved experimentally \cite{Ronen2018,Wu2018}. The device
layout is also adaptable to other PF platforms, in particular to the
FQH case $\nu=2/3$ \cite{Mong2014}. The theory recovers the Majorana-based
TKE \cite{Beri2012,Altland2013,Beri2013} for $\nu=1$.

As shown in Fig.~\ref{fig1}, several additional FQH edges can now
serve as probing leads in transport studies. Fractional quasiparticles
in these edges are thereby tunnel-coupled to the PF operator $\hat{\alpha}_{j}$
at the respective domain wall. A general setup consists of a PF box
made of $N$ SC domains and contacted by $3\leq M\leq2N$ edge states.
We show the simplest non-trivial case with $N=2$ and $M=3$ in Fig.~\ref{fig1}.
It is important that the PF box device is kept floating (not grounded)
such that the total charge on the PF box is restricted by the Coulomb
charging energy.

Let us first outline the theoretical description of the FQH edge state
leads. Each of the $M$ lead pieces is described by a chiral boson
field, $\hat{\phi}_{j}(x)$, with the Hamiltonian \cite{Wen1991,Kane1992a,Gogolin1998,Altland2010}
\begin{equation}
H_{\mathrm{edge}}=\sum_{j=1}^{M}\frac{v}{4\pi}\int_{-L/2}^{+L/2}dx(\partial_{x}\hat{\phi}_{j})^{2},\label{eq:H_edge}
\end{equation}
where $v$ is the edge velocity, assumed identical in all leads. Anisotropies
in these velocities do not cause physical effects since they can be
absorbed by a renormalization of cotunneling amplitudes. Since we
are not interested in finite-size effects in the leads, we will also
assume $L\to\infty$. The commutation relations between chiral boson
fields, 
\begin{equation}
[\hat{\phi}_{j}(x),\hat{\phi}_{k}(x')]=i\pi\left[\delta_{jk}\sgn(x-x')+\sgn(k-j)\right],\label{phicomm}
\end{equation}
already incorporate Klein factors \cite{Gogolin1998} since Eq.~\eqref{phicomm}
follows from a single-edge commutation relation by imagining that
all leads actually belong to one long edge. It is important, however,
that different leads are dynamically independent, which in turn is
ensured by the Ohmic contacts in Fig.~\ref{fig1}. The fractional
quasiparticle operator can then be expressed as vertex operator of
the respective chiral boson field, $\hat{\psi}_{j}(x)\sim\ e^{i\sqrt{\nu}\hat{\phi}_{j}(x)}$,
see~Refs.~\cite{Wen1991,Kane1992a,Gogolin1998,Altland2010} and
App.~\ref{appA}.

Next, we summarize the theoretical description of the PF box. For
detailed derivations and discussions, see Refs.~\cite{Lindner2012,Clarke2013,Snizhko2018}.
The PF box is defined from opposite-spin FQH edges which are proximitized
by alternating FM and SC domains, see Fig.~\ref{fig1}. Since at
low energy scales, these domains are gapped, operators creating low-energy
excitations can only reside at the domain walls in between adjacent
domains. Similar to the Majorana case, the domain wall hosts stable
zero-energy modes corresponding to the PF operators $\hat{\alpha}_{j}$.
The latter obey the $\mathbb{Z}_{n}$ PF algebra with index $n=2/\nu$,
\begin{equation}
\hat{\alpha}_{j}\hat{\alpha}_{k}=\omega^{\sgn(k-j)}\hat{\alpha}_{k}\hat{\alpha}_{j},\quad\omega=e^{2\pi i/n}=e^{i\pi\nu}.\label{PFalgebra}
\end{equation}
The low-energy PF box Hilbert space is spanned by the states $|Q_{\mathrm{tot}},Q_{1}\mathrm{mod}~2,...,Q_{N-1}\mathrm{mod}~2\rangle$,
where $Q_{{\rm tot}}$ is the total charge of the proximitizing SCs
and the FQH edges within the PF box. Importantly, $Q_{{\rm tot}}$
has fractional values differing by multiples of $\nu$. We note that
here all SC$_{j}$ pieces are implied to be part of one floating superconductor,
and similarly all FM$_{j}$ are part of one ferromagnet. The quantum
numbers $Q_{j}$ describe the charge of the FQH edges trapped between
FM$_{j}$ and FM$_{j+1}$, and are also quantized in units of $\nu$.
Since the proximitizing SCs can absorb Cooper pairs at no kinetic
energy cost, the $Q_{j}$ are only defined modulo 2. These quantum
numbers correspond to the distribution of fractional quasiparticles
between different SC$_{j}$ parts. The box Hamiltonian only receives
a Coulomb charging contribution sensitive to the total charge $Q_{{\rm tot}}$,
\begin{equation}
H_{\mathrm{box}}=E_{C}\left(\hat{Q}_{{\rm tot}}-q_{0}\right)^{2}.\label{eq:H_box}
\end{equation}
By changing a backgate voltage, one can tune the parameter $q_{0}$
such that the box has quantized ground-state charge given by the value
of $Q_{{\rm tot}}$ closest to $q_{0}$. For special choices of $q_{0}$,
one may also reach charge degenerate points, but such cases are not
considered here.

Finally, we include complex-valued tunneling amplitudes $\eta_{j}$
describing tunneling of quasiparticles between the respective edge
($\hat{\psi}_{j}$) through the FQH bulk to the PF box via $\hat{\alpha}_{j}$.
Assuming a point-like tunnel contact at $x=0$ along the respective
edge, the tunneling Hamiltonian is given by \cite{Lindner2012,Snizhko2018}
\begin{equation}
H_{\mathrm{tun}}=\sum_{j=1}^{M}\eta_{j}\hat{\psi}_{j}(0)\hat{\alpha}_{j}^{\dagger}+\mathrm{h.c.}\label{htun11}
\end{equation}
Using Eq.~\eqref{phicomm}, the fractional quasiparticle operators
$\hat{\psi}_{j}(x)\sim e^{i\sqrt{\nu}\hat{\phi}_{j}(x)}$ obey the
algebra 
\begin{equation}
\hat{\psi}_{j}(x)\hat{\psi}_{k}(x')=e^{-i\pi\nu\sgn(k-j)-i\pi\nu\delta_{jk}\sgn(x-x')}\hat{\psi}_{k}(x')\hat{\psi}_{j}(x).\label{psicomm}
\end{equation}
Furthermore, one can show that $\left[\hat{\phi}_{j}(x),\hat{\alpha}_{k}\right]=-\pi\sqrt{\nu}$
holds. From the latter relation, we obtain 
\begin{equation}
\hat{\psi}_{j}(x)\hat{\alpha}_{k}=e^{-i\pi\nu}\hat{\alpha}_{k}\hat{\psi}_{j}(x).\label{eq:PFqp_permutation}
\end{equation}
Altogether these relations imply that all terms in Eq.~\eqref{htun11}
commute with each other. This fact will become important when we discuss
the strong-coupling regime in Sec.~\ref{sec4}. We emphasize that
Klein factors, which are needed to ensure proper statistical phase
relations between different edges \cite{Gogolin1998}, are fully taken
into account by Eqs.~\eqref{phicomm}, \eqref{PFalgebra}, \eqref{psicomm}
and \eqref{eq:PFqp_permutation}.

For a generic backgate parameter $q_{0}$ in Eq.~\eqref{eq:H_box},
the ground state of the PF box has quantized charge $Q_{\mathrm{tot}}$
due to the large charging energy $E_{C}$. The dominant low-energy
processes then come from the cotunneling of fractional quasiparticles
between different edges mediated by the PF box. Technically, one obtains
the cotunneling Hamiltonian, $H_{{\rm cot}}$, by projecting the full
Hamiltonian, $H=H_{{\rm edge}}+H_{{\rm box}}+H_{{\rm tun}}$, to the
charge ground-state sector of the PF box, $H\to H_{{\rm eff}}=H_{{\rm edge}}+H_{{\rm cot}}.$
A standard Schrieffer-Wolff transformation \cite{Altland2010,Schrieffer1966}
yields 
\begin{eqnarray}
H_{\mathrm{cot}} & = & -\sum_{j,k=1;j\neq k}^{M}\lambda_{jk}\hat{\psi}_{j}^{\dagger}(0)\hat{\psi}_{k}^ {}(0)\hat{\alpha}_{j}^ {}\hat{\alpha}_{k}^{\dagger}\label{eq:H_cot}\\
 & - & \sum_{j}\left|\eta_{j}\right|^{2}\left(U_{+}^{-1}\hat{\psi}_{j}^{\dagger}(0)\hat{\psi}_{j}^ {}(0)+U_{-}^{-1}\hat{\psi}_{j}^ {}(0)\hat{\psi}_{j}^{\dagger}(0)\right),\nonumber 
\end{eqnarray}
where the cotunneling amplitude from lead $k$ to lead $j$ is 
\begin{equation}
\lambda_{jk}=\eta_{j}^{*}\eta_{k}^ {}\left(U_{+}^{-1}+U_{-}^{-1}\right).\label{cotunnel}
\end{equation}
Here, $U_{+}$ ($U_{-})$ denotes the energy cost for adding (removing)
one fractional quasiparticle to (from) the box. For instance, assuming
$|q_{0}|<\nu/2$, one finds $U_{\pm}=E_{C}\nu^{2}(1\mp2q_{0}/\nu)$.
As shown in App.~\ref{appA}, for $\nu\le1$, the potential scattering
terms corresponding to the second row in Eq.~\eqref{eq:H_cot} can
always be neglected. In addition, the complex phases of $\eta_{j}$
can be gauged away by shifting the respective boson field, $\hat{\phi}_{j}(x)\to\hat{\phi}_{j}(x)+{\rm cst}$.
This gauge transformation renders all $\lambda_{jk}$ in Eq.~\eqref{cotunnel}
real positive and symmetric, $\lambda_{kj}=\lambda_{jk}>0$. We note
in passing that the total electric charge on the PF box is explicitly
preserved by Eq.~\eqref{eq:H_cot}, as well as the total $\mathbb{Z}_{n}$
charge described in Ref.~\cite{Alicea2016}. In addition, we remark
that the original derivation by Schrieffer and Wolff contains a term
beyond Eq.~\eqref{eq:H_cot}, see Eq.~(12) in Ref.~\cite{Schrieffer1966}.
A similar term describing the simultaneous tunneling of two quasi-particles
onto/off the impurity arises in our case. However, this term does
not preserve the PF box electric charge and thus vanishes after the
projection to the charge ground state.

Our PF generalization of the topological Kondo model thus corresponds
to the effective low-energy Hamiltonian 
\begin{equation}
H_{{\rm eff}}=\sum_{j=1}^{M}\frac{v}{4\pi}\int_{-\infty}^{\infty}dx(\partial_{x}\hat{\phi}_{j})^{2}-\sum_{j\neq k}^{M}\lambda_{jk}\hat{\psi}_{j}^{\dagger}(0)\hat{\psi}_{k}^ {}(0)\hat{\alpha}_{j}\hat{\alpha}_{k}^{\dagger},\label{PFK}
\end{equation}
together with the commutation relations in Eqs.~\eqref{phicomm},
\eqref{PFalgebra}, \eqref{psicomm}, and \eqref{eq:PFqp_permutation}.

At this point several remarks are in order. 

\begin{figure}
\begin{centering}
\includegraphics[width=0.45\textwidth]{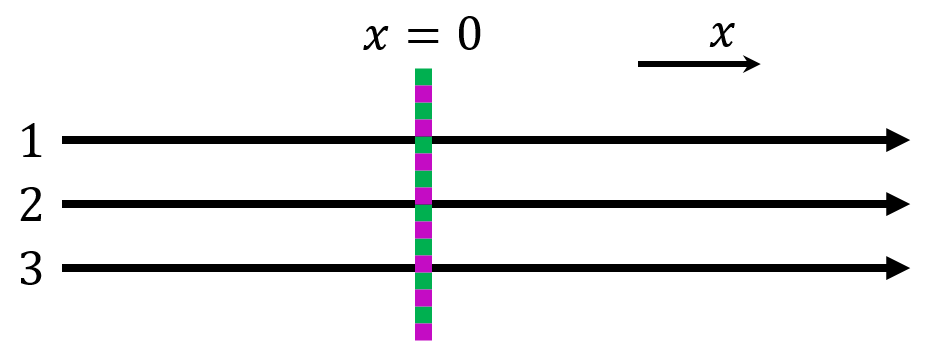} 
\par\end{centering}
\caption{Schematic view of the effective Hamiltonian $H_{{\rm eff}}$ in Eq.~\eqref{PFK},
which defines a PF generalization of the topological Kondo model.
Here, $M=3$ fractional edge state leads are connected via pairwise
cotunneling processes at $x=0$. Each cotunneling event comes with
a transition in the PF Hilbert space (indicated by the vertical colored
bar). }
\label{fig2} 
\end{figure}

{[}i){]} 
\begin{enumerate}
\item For $\nu=1$, noting that $\hat{\alpha}_{j}\hat{\alpha}_{k}^{\dagger}\to\gamma_{j}\gamma_{k}$,
the quantum impurity spin operator in Eq.~\eqref{PFK} has the components
$i\gamma_{j}\gamma_{k}$. These Majorana bilinears generate the algebra
so$(M)$ \cite{Beri2012,Zazunov2014}, and Eq.~\eqref{PFK} reduces
to the TKE Hamiltonian. For $M=3$, the three independent bilinears
are equivalently expressed by standard Pauli operators, so$(3)=$~su$(2)$. 
\item For $\nu<1$, the PF box also has a continuous symmetry. The PF bilinears
$\hat{\alpha}_{j}\hat{\alpha}_{k}^{\dagger}$ appearing in Eq.~\eqref{PFK}
do not constitute a closed Lie algebra. However, together with their
powers and products of those, they close the algebra su$\left(n^{[(M-1)/2]}\right)$,
where $[x]$ is the integer part of $x$, acting onto the PF box Hilbert
space in the fundamental representation \cite{footnote1}. In particular,
for $M=3$, the algebra su$(n)$ is generated by the set of operators
\begin{equation}
\left\{ \hat{\alpha}_{1}^{k_{1}}\hat{\alpha}_{2}^{k_{2}}\hat{\alpha}_{3}^{-k_{1}-k_{2}}\right\} ,\label{adjoint}
\end{equation}
where integers $k_{j}$ are defined modulo $n=2/\nu$ as $\hat{\alpha}_{j}^{n}=1$,
and we use the convention $\hat{\alpha}_{j}^{-k}\equiv(\hat{\alpha}_{j}^{\dagger})^{k}$
for $k>0$. The fact that the dimension of the `quantum impurity'
representation space is $n$ follows from the PF representation in
Refs.~\cite{Alicea2016,Dong2017} together with the $\mathbb{Z}_{n}$-charge
conservation constraint. For $\nu=1/3,$ we arrive at su(6) with its
35 generators plus the identity, which are given by Eq.~\eqref{adjoint}.
Note that the bilinears $\hat{\alpha}_{j}\hat{\alpha}_{k}^{\dagger}$
appearing in the Hamiltonian \eqref{PFK} themselves constitute only
a small subset of six out of the 35 algebra generators. 
\item The leads, however, in general do not possess a continuous symmetry,
cf.~App.~\ref{appb}. This situation should be contrasted to standard
Kondo problems (including the TKE), where both the impurity and the
leads constitute representations of a symmetry group, and the interaction
between them is built out of currents generating this symmetry in
each part. 
\item Nonetheless, Eq.~\eqref{PFK} shows that basic ingredients of a typical
quantum impurity setting are present. A schematic sketch of $H_{{\rm eff}}$
in Eq.~\eqref{PFK} is depicted in Fig.~\ref{fig2}, where $M$
(parallel) chiral edges interact by cotunneling processes of fractional
quasiparticles at $x=0$ between different lead pairs. Simultaneously,
such exchange processes cause transitions in the PF box Hilbert space
via the PF bilinears $\sim\hat{\alpha}_{j}\hat{\alpha}_{k}^{\dagger}$. 
\end{enumerate}

\section{RG analysis}

\label{sec3}

We now turn to the weak-coupling regime and discuss the one-loop RG
equations for the PF generalization of the topological Kondo model
in Eq.~\eqref{PFK}. To that end, consider a perturbative expansion
of the partition sum in powers of the cotunneling amplitudes $\lambda_{jk}=\lambda_{kj}>0$.
Within the RG approach \cite{Altland2010}, upon reducing the effective
bandwidth $\Lambda$ from its bare value, $\Lambda(\ell=0)\approx E_{C}$,
one analyzes how these couplings are renormalized and whether new
types of couplings will be generated. Writing $\Lambda(\ell)=\Lambda(0)e^{-\ell}$,
the scale $\Lambda(\ell)$ refers to the energy scale at which the
system is probed with increasing RG flow parameter $\ell$. In case
the RG flow of the $\lambda_{jk}(\ell)$ approaches the strong-coupling
limit, the RG approach breaks down at $\ell=\ell^{*}$ where one hits
a divergence of the dominant coupling. The corresponding scale defines
the Kondo temperature, 
\begin{equation}
T_{K}=\Lambda(\ell^{*})=E_{C}e^{-\ell^{*}}.\label{TK}
\end{equation}
The physics in the strong-coupling regime, i.e., for energies well
below $T_{K}$, will be addressed in Sec.~\ref{sec4}.

The one-loop RG equations are most conveniently obtained via the operator
product expansion (OPE) approach \cite{Cardy1996}, where one considers
arbitrary pairs of cotunneling operators in Eq.~\eqref{PFK} at almost
coinciding times $t$ and $t'$. The result of such a contraction
must be equivalent to a linear combination of all possible boundary
operators taken at time $(t+t')/2$, and the expansion coefficients
directly determine the one-loop RG equations \cite{Cardy1996}. One
thus has to analyze contractions of cotunneling operators, cf.~App.~\ref{appA}.
Most contractions imply a renormalization of a cotunneling amplitude
$\lambda_{jk}$. However, one also finds additional contributions
generating new couplings. For instance, a contraction of the cotunneling
operators corresponding to $\lambda_{kj}$ and $\lambda_{km}$ (with
$k\ne j\ne m$) generates a coupling to the operator 
\begin{equation}
H'=\left(\hat{\psi}_{k}^{\dagger}(0)\right)^{2}\hat{\psi}_{j}(0)\hat{\psi}_{m}(0)\hat{\alpha}_{k}^{2}\hat{\alpha}_{j}^{\dagger}\hat{\alpha}_{m}^{\dagger},\label{newop}
\end{equation}
which has scaling dimension $\Delta'=3\nu$. Since the cotunneling
terms $\sim\lambda_{jk}$ in Eq.~\eqref{PFK} have the smaller scaling
dimension $\Delta_{jk}=\nu$, we conclude that $H'$ is much less
relevant (in fact marginal for $\nu=1/3$, and irrelevant for $\nu=2/3$)
and thus can be dropped. Contractions of cotunneling operators then
only produce PF bilinears of the type $\sim\hat{\alpha}_{j}\hat{\alpha}_{k}^{\dagger}$
already present in Eq.~\eqref{PFK}. In particular, for $M=3$, all
other impurity operators in the set \eqref{adjoint} are not activated.

Remarkably, after a uniform rescaling, 
\begin{equation}
\tilde{\lambda}_{jk}=\frac{2\tau_{c}^{1-\nu}}{v^{\nu}}\lambda_{jk},
\end{equation}
where $\tau_{c}$ denotes a short-time cutoff, we find that the RG
equations for our PF generalization \emph{coincide} with those for
the Majorana-based TKE with Luttinger liquid parameter $g=1/\nu$
\cite{Altland2013,Beri2013}, 
\begin{equation}
\frac{d\tilde{\lambda}_{jk}}{d\ell}=(1-\nu)\tilde{\lambda}_{jk}+\sum_{m\neq(j\ne k)}^{M}\tilde{\lambda}_{jm}\tilde{\lambda}_{mk}.\label{RGeq}
\end{equation}
Consider first the isotropic part of the cotunneling couplings. Writing
$\tilde{\lambda}_{jk}=\tilde{\lambda}(1-\delta_{jk})$, Eq.~\eqref{RGeq}
yields 
\begin{equation}
\frac{d\tilde{\lambda}}{d\ell}=(1-\nu)\tilde{\lambda}+(M-2)\tilde{\lambda}^{2},
\end{equation}
which is solved by 
\begin{equation}
\tilde{\lambda}(\ell)=\frac{\tilde{\lambda}(0)e^{(1-\nu)\ell}}{1+\frac{M-2}{1-\nu}\tilde{\lambda}(0)[1-e^{(1-\nu)\ell}]}.
\end{equation}
Clearly, the isotropic part $\tilde{\lambda}(\ell)$ flows towards
strong coupling and diverges once the running energy scale reaches
the Kondo temperature in Eq.~\eqref{TK}. We find 
\begin{equation}
T_{K}\simeq E_{C}\left(\frac{(M-2)\tilde{\lambda}(0)}{1-\nu}\right)^{1/(1-\nu)}.\label{tkd}
\end{equation}
The power-law dependence of $T_{K}$ on the average cotunneling coupling
$\tilde{\lambda}(0)$ should be contrasted to the $T_{K}\sim e^{-1/[(M-2)\tilde{\lambda}(0)]}$
law of the TKE for $\nu=1$ \cite{Beri2012}. Equation \eqref{tkd}
thus suggests that much higher $T_{K}$ are possible for $\nu<1$.
We note that although the notation $T_{K}$ is suggestive of a `Kondo
temperature', Eq.~\eqref{tkd} only indicates the separation between
the regimes of strong and weak coupling. Indeed, as discussed above,
for $\nu<1$, the leads do not possess a continuous symmetry, and
hence one cannot speak of Kondo screening processes in the usual sense.
Finally, in order to show that anisotropies in the cotunneling amplitudes
are negligible as in the conventional TKE \cite{Beri2012}, we have
linearized the RG equations \eqref{RGeq} in the $\tilde{\lambda}_{jk}$
anisotropies taken relative to the isotropic component. Our analysis
shows that relative anisotropies are RG irrelevant and thus can be
neglected at low energy scales.

\section{Strong-coupling solution}

\label{sec4}

We next turn to the strong-coupling regime realized at energy scales
well below $T_{K}$ in Eq.~\eqref{tkd}. We here pursue a similar
strategy as done for the TKE in Refs.~\cite{Altland2013,Beri2013,Zazunov2014}.
To that end, we take the bosonized version of Eq.~\eqref{PFK}, with
$\hat{\psi}_{j}(0)\sim e^{i\sqrt{\nu}\hat{\phi}_{j}(0)}$, and consider
the limit $\lambda_{jk}\to+\infty$. As a result, in the ground state
of the system, the phase fields $\hat{\phi}_{j}(0)$ will be locked
into a configuration minimizing the cosine potentials resulting from
$H_{{\rm cot}}$. Using the general approach of Ref.~\cite{Ganeshan2016},
one can then construct the low-energy Hamiltonian describing this
fixed point. We also systematically compute all possible operator
perturbations around the fixed point and thereby show that it is stable.
The leading irrelevant operators then determine the finite-$T$ corrections
to the conductance tensor, where we will compare our results to the
TKE expression in Eq.~\eqref{tkecond}.

\subsection{Strong-coupling fixed point}

\label{sec4a}

Using Eqs.~\eqref{PFalgebra}, \eqref{psicomm} and \eqref{eq:PFqp_permutation},
one finds that all cotunneling operators $\hat{\psi}_{j}^{\dagger}(0)\hat{\psi}_{k}(0)\hat{\alpha}_{j}\hat{\alpha}_{k}^{\dagger}$
in Eq.~\eqref{PFK} are mutually commuting and thus can be diagonalized
simultaneously. At the putative strong-coupling fixed point, we now
demand that each of those terms is separately minimized. Defining
auxiliary operators $\hat{\beta}_{jk}$, 
\begin{equation}
\hat{\alpha}_{j}\hat{\alpha}_{k}^{\dagger}=e^{i\pi\hat{\beta}_{jk}-i\pi\nu\sgn(k-j)},
\end{equation}
Equation~\eqref{PFK} yields 
\begin{equation}
H_{{\rm cot}}=-\frac{2}{(v\tau_{c})^{\nu}}\sum_{j>k}\lambda_{jk}\cos\left(\sqrt{\nu}\left[\hat{\phi}_{k}(0)-\hat{\phi}_{j}(0)\right]+\pi\hat{\beta}_{jk}\right).\label{eq:H_PF_Kondo_cosines}
\end{equation}
Minimizing $H_{{\rm cot}}$ for $\lambda_{jk}\to+\infty$ is then
equivalent to imposing the constraints 
\begin{equation}
\sqrt{\nu}\left[\hat{\phi}_{k}(0)-\hat{\phi}_{j}(0)\right]+\pi\hat{\beta}_{jk}=2\pi\hat{C}_{jk},\label{constraint}
\end{equation}
where $\hat{C}_{jk}$ are integer-valued operators. We note that the
commutation relations $[\hat{\beta}_{jk},\hat{\phi}_{m}(x)]=0$ and
\begin{eqnarray}
[\hat{\beta}_{jk},\hat{\beta}_{mn}] & = & i\frac{\nu}{\pi}\Bigl[\sgn(j-m)+\sgn(k-n)\nonumber \\
 & - & \sgn(j-n)-\sgn(k-m)\Bigr]
\end{eqnarray}
imply that $[\hat{C}_{jk},\hat{C}_{mn}]=0.$ In addition, with $\hat{\beta}_{jk}+\hat{\beta}_{km}=\hat{\beta}_{jm}$,
we see that $\hat{C}_{jk}+\hat{C}_{km}=\hat{C}_{jm}$. As a consequence,
to enforce that all $\hat{C}_{jk}$ are integer, it suffices to demand
that all $\hat{C}_{1j}$ with $j=2,\ldots,M$ are integer-valued operators.

For constructing the low-energy theory, we next employ the powerful
approach of Ref.~\cite{Ganeshan2016}, which is tailor-made to solving
problems with large-amplitude cosine potentials as encountered here.
According to this method, the $\hat{C}_{jk}$ should be constrained
to being integer numbers, where the low-energy Hamiltonian $H_{\mathrm{LE}}$
follows from the free part, $H_{0}=H_{\mathrm{edge}}$ in Eq.~\eqref{eq:H_edge},
minus all terms causing a non-trivial time evolution of the $\hat{C}_{jk}$
in Eq.~\eqref{constraint}. The resulting Hamiltonian is quadratic
and can easily be quantized. Delegating technical details to App.~\ref{appc},
the quantized phase fields $\hat{\phi}_{j}(x,t)$ are expressed in
terms of standard boson operators $\hat{a}_{q,j}$ (zero-mode operators
$\hat{\phi}_{0,j}$) for each $\omega\neq0$ ($\omega=0$) eigenfunction,
cf.~Eq.~\eqref{eq:phi_j_quantized} in App.~\ref{appA}. To that
end, we define the matrix 
\begin{equation}
U_{jk}=\frac{2}{M}-\delta_{jk},\label{udef}
\end{equation}
and the operators 
\begin{eqnarray}
\hat{f}_{j}(x,t) & = & \hat{\phi}_{0,j}+\hat{g}_{j}(x,t),\nonumber \\
\hat{g}_{j}(x,t) & = & i\int_{0}^{\infty}\frac{dq}{\sqrt{q}}\left(\hat{a}_{q,j}e^{iq(x-vt)}-{\rm h.c.}\right),\label{gdef}
\end{eqnarray}
with the commutation relations 
\begin{equation}
\left[\hat{a}_{q,j},\hat{a}_{q',m}^{\dagger}\right]=\delta(q-q')\delta_{jm},\quad\left[\hat{a}_{q,j},\hat{a}_{q',m}\right]=0,
\end{equation}
and $\left[\hat{\phi}_{0,j},\hat{\phi}_{0,m}\right]=i\pi\sgn(m-j)$.
For incoming states ($x<0$), the chiral boson field is 
\begin{equation}
\hat{\phi}_{j}(x<0,t)=\hat{f}_{j}(x,t),\label{phiii}
\end{equation}
while outgoing states ($x>0$) are given by 
\begin{eqnarray}
\hat{\phi}_{j}(x>0,t) & = & \sum_{k=1}^{M}U_{jk}\hat{f}_{k}(x,t)+\label{phii}\\
 & + & \frac{2\pi}{M}\sqrt{\nu^{-1}}\sum_{k=1,k\neq j}^{M}\left(\hat{\beta}_{jk}-2\hat{C}_{jk}\right).\nonumber 
\end{eqnarray}
The field at $x=0$ follows from the above relations, $\hat{\phi}_{j}(0,t)=[\hat{\phi}_{j}(0^{+},t)+\hat{\phi}_{j}(0^{-},t)]/2$,
and the low-energy Hamiltonian $H_{{\rm LE}}$ takes the form, cf.~Eq.~\eqref{eq:H_LE}
in App.~\ref{appc}, 
\begin{equation}
H_{\mathrm{LE}}=\sum_{j=1}^{M}\int_{0}^{\infty}dq\ vq\left(\hat{a}_{q,j}^{\dagger}\hat{a}_{q,j}+\frac{1}{2}\right).
\end{equation}

For a discussion of transport features, we next note that the current
flowing along an edge corresponds to the operator \cite{Wen1991,Kane1992a,Gogolin1998,Altland2010}
\begin{equation}
\hat{I}_{j}(x,t)=-\frac{\sqrt{\nu}}{2\pi}\partial_{t}\hat{\phi}_{j}(x,t).
\end{equation}
Using Eqs.~\eqref{udef}\textendash \eqref{phii}, we find 
\begin{equation}
\hat{I}_{j}(x,t)=-\frac{\sqrt{\nu}}{2\pi}\sum_{k}\left[\theta(x)U_{jk}+\theta(-x)\delta_{jk}\right]\partial_{t}\hat{g}_{k}(x,t),
\end{equation}
with the Heaviside function $\theta(x)$. We thus observe that 
\begin{equation}
\hat{I}_{j}(0^{+},t)=\sum_{k=1}^{M}U_{jk}\hat{I}_{k}(0^{-},t).
\end{equation}
At low frequencies $\omega=vq\to0$, we obtain 
\begin{equation}
I_{j}^{+}=\sum_{k}U_{jk}I_{k}^{-}=\nu\frac{e^{2}}{h}\sum_{k}U_{jk}V_{k},
\end{equation}
where $I_{j}^{+/-}=\left\langle \hat{I}_{j}(x>0/x<0,t)\right\rangle $
refers to the scattered/incoming current, respectively. The incoming
current is given by $I_{j}^{-}=\nu e^{2}V_{j}/h$, where $V_{j}$
is the voltage for injected quasiparticles at the $j$th edge.

At the strong-coupling fixed point ($T=V_{j}=0$), we thus obtain
the universal multi-terminal conductance tensor 
\begin{equation}
G_{jk}=\frac{dI_{j}^{+}}{dV_{k}}=\frac{\nu e^{2}}{h}U_{jk}=\frac{\nu e^{2}}{h}\left(\frac{2}{M}-\delta_{jk}\right).\label{t0cond}
\end{equation}
For $\nu=1$, taking into account that the injection and collection
points are spatially separated in our Hall setup, Eq.~\eqref{t0cond}
has the same physical content as the $T=0$ TKE conductance in Eq.~\eqref{tkecond}.
Indeed, Eq.~\eqref{t0cond} describes the scattered current $I_{j}^{+}$.
Studying instead the current at the tunnel contact, $I_{j}=I_{j}^{+}-I_{j}^{-}$,
we have to replace $U_{jk}\to U_{jk}-\delta_{jk}=2\left(\frac{1}{M}-\delta_{jk}\right)$
in Eq.~\eqref{t0cond}. After this step, Eq.~\eqref{t0cond} matches
the TKE conductance tensor in Eq.~\eqref{tkecond} taken at $T=0$.
Remarkably, the isotropic structure of the TKE conductance \eqref{tkecond}
carries over to the $\nu<1$ PF generalization, despite of the fact
that we are not dealing with a Kondo problem anymore.

A particularly noteworthy consequence of Eq.~\eqref{t0cond} is revealed
by inspecting the diagonal component of the conductance tensor, which
has the universal, fractionally quantized, and negative value 
\begin{equation}
G_{jj}=-\frac{M-2}{M}\frac{\nu e^{2}}{h}<0.
\end{equation}
Since this conductance is negative, our device can be operated as
\emph{current extractor}. For instance, putting all $V_{k}=0$ except
for $V_{1}\ne0$, current is injected only via the first lead, $I_{1}^{-}=\frac{\nu e^{2}}{h}V_{1}\equiv I_{{\rm in}}$.
The outgoing current in this lead then has the opposite sign, $I_{1}^{+}=-\frac{M-2}{M}I_{{\rm in}}<0$,
where the fraction $-I_{1}^{+}/I_{{\rm in}}$ is determined only by
the number $M$ of leads. For this example, the outgoing currents
in all other $M-1$ leads ($j>1$) are identical and given by $I_{j}^{+}=\frac{2}{M}I_{{\rm in}}$,
see Eq.~\eqref{t0cond}. Current conservation, $I_{{\rm out}}=(M-1)I_{j}^{+}+I_{1}^{+}=I_{{\rm in}}$,
thus requires that current must be extracted from lead $j=1$. Similar
current extraction phenomena in quantum Hall devices have been discussed
in Ref.~\cite{Protopopov2017}.

Another striking consequence of Eq.~\eqref{t0cond} is the absence
of current-current correlations between different terminals at the
fixed point, which can be established from the above theory along
the lines of Refs.~\cite{Beri2013,Zazunov2014}. The absence of shot
noise is noteworthy since incoming currents are partitioned into currents
flowing through all leads attached to the PF box, see Eq.~\eqref{t0cond},
and such partitioning processes usually generate noise \cite{Altland2010}.
Noiseless partitioning of currents in multi-terminal quantum junctions
has also been established for the TKE \cite{Beri2013,Zazunov2014}
and for the related case of quantum Hall junctions coupled through
a central quantum dot \cite{Altland2012,Altland2015}. In our system,
leading irrelevant operators, see Sec.~\ref{sec4b}, can be responsible
for weak contributions to shot noise. However, such contributions
quickly vanish as one approaches the fixed point for $T,V_{j}\to0$.

\subsection{Stability of the strong-coupling point}

\label{sec4b}

The stability of the strong-coupling fixed point and the low-energy
physics in its vicinity are determined by the leading irrelevant operators
(LIOs), where we anticipate that our analysis finds no marginal or
relevant operators perturbing the fixed point. Such perturbations
will appear because the $\lambda_{jk}$ couplings are large but finite
and are constructed from admissible operators at the fixed point.
The latter have to obey three requirements, namely (i) they do not
change the charge $Q_{\mathrm{tot}}$ of the PF box, (ii) they change
the total charge of each edge only in multiples of $\nu$, and (iii)
they alter $\hat{C}_{jk}$ in Eq.~\eqref{constraint} only by an
integer number. Condition (i) prohibits operators with an odd number
of PF operators $\hat{\alpha}_{m}^{\thinspace},\hat{\alpha}_{m}^{\dagger}$.
Condition (ii) implies that operators $e^{\pm i\sqrt{\nu}\hat{\phi}_{j}(x,t)}$
(or multiples thereof) are involved. Finally, condition (iii) further
constrains the set of allowed operators. By using (iii) in conjunction
with the commutation relations 
\begin{multline}
\left[\hat{C}_{jk}(t),e^{\pm i\sqrt{\nu}\hat{\phi}_{m}(x,t)}\right]=\pm\frac{\nu}{2}\Bigl(\sgn(x)\left(\delta_{km}-\delta_{jm}\right)\\
+\sgn(k-m)-\sgn(j-m)\Bigr)e^{\pm i\sqrt{\nu}\hat{\phi}_{m}(x,t)}\label{eq:constraints_qps_commutation}
\end{multline}
and $\left[\hat{C}_{jk}(t),e^{i\sqrt{\nu}\hat{\phi}_{m}(0,t)}\hat{\alpha}_{m}^{\dagger}\right]=0$,
one can determine all admitted operators at the strong-coupling fixed
point. The list of elementary allowed operators is given in Table~\ref{tab1}.
All non-trivial allowed operators can be constructed by taking composites
of the operators in Table~\ref{tab1}. Note that the list implies
that electrons can tunnel in and out of the system anywhere at $x\neq0$,
but quasiparticles can only tunnel outside of the whole structure
or between neighboring edges. Since quasiparticles have to tunnel
through the FQH bulk, these are also the only physical possibilities
in Fig.~\ref{fig1}.

\begin{table}
\noindent \begin{centering}
\begin{tabular}{|c|>{\centering}p{0.5\columnwidth}|}
\hline 
$\hat{O}$  & $\Delta_{\hat{O}}$\tabularnewline
\hline 
\hline 
$e^{\pm i\sqrt{\nu}\hat{\phi}_{1}(x>0,t)}$  & $\nu/2$\tabularnewline
\hline 
$e^{\pm i\sqrt{\nu}\hat{\phi}_{M}(x<0,t)}$  & $\nu/2$\tabularnewline
\hline 
$e^{\pm i\sqrt{\nu}\left(\hat{\phi}_{j}(x>0,t)-\hat{\phi}_{j-1}(x'<0,t)\right)}$  & $\nu\left(1-\frac{2}{M}\right)$,\linebreak{}
 when $x=-x'=0^{+}$\tabularnewline
\hline 
$e^{\pm i\sqrt{\nu^{-1}}\hat{\phi}_{j}(x\neq0,t)}$  & $\nu^{-1}/2$\tabularnewline
\hline 
\end{tabular}
\par\end{centering}
\caption{\label{tab1} Elementary allowed operators $\hat{O}$ and their scaling
dimensions $\Delta_{\hat{O}}$ at the strong-coupling fixed point.
Combinations of them lead to non-trivial new scaling dimensions.}
\end{table}

The scaling dimensions of operators and their combinations are easily
obtained by expressing operators in terms of $\hat{g}_{j}(x,t)$,
see Eq.~\eqref{gdef}, and ignoring zero modes. Indeed, the operator
$e^{i\sum_{j}p_{j}\hat{g}_{j}(x,t)}$ is seen to have scaling dimension
$\Delta=\sum_{j}p_{j}^{2}/2$ by means of the relation 
\begin{equation}
\left\langle e^{i\sum_{j}p_{j}\hat{g}_{j}(x,t)}e^{-i\sum_{j}p_{j}\hat{g}_{j}(x,t')}\right\rangle \sim(t-t'-i0^{+})^{-\sum_{j}p_{j}^{2}}.\label{eq:exp_scal_dim}
\end{equation}
Next we calculate the scaling dimension of the LIO. Within the original
Hamiltonian \eqref{PFK}, transfer of charge between different edges
is only possible through exponentials of $\hat{C}_{jk}$. A non-trivial
perturbation to the strong-coupling fixed point can then only result
from edge fields $\hat{\phi}_{m}(x,t)$ taken near $x=0$ on a \emph{single}
edge $m$. The most general perturbation has the form 
\begin{equation}
\hat{O}_{c}\sim e^{i\sqrt{\nu^{-1}}\left(d_{+}\hat{\phi}_{m}(0^{+},t)+d_{-}\hat{\phi}_{m}(0^{-},t)\right)},\quad d_{\pm}\in\mathbb{Z}.
\end{equation}
Since this operator should conserve total charge, we require $d_{-}=-d_{+}$.
Then $\hat{O}_{c}$ shifts $\hat{C}_{jk}\rightarrow\hat{C}_{jk}+d_{+}\left(\delta_{km}-\delta_{jm}\right)$,
while its scaling dimension is given by $\Delta_{c}=d_{+}^{2}\frac{2}{\nu}\left(1-\frac{1}{M}\right)$.
The LIO follows for $d_{+}=\pm1$ (where $\hat{O}_{c}$ coincides
with $e^{\mp i\hat{\Pi}_{m}}$ in App.~\ref{appc}), with the scaling
dimension 
\begin{equation}
\Delta_{\mathrm{LIO}}=\frac{2}{\nu}\left(1-\frac{1}{M}\right).
\end{equation}
For all $M\geq3$ and $\nu\leq1$, we observe that $\Delta_{\mathrm{LIO}}>1$.
The fixed point is thus stable as asserted before. Furthermore, since
$\Delta_{\mathrm{LIO}}=\Delta_{M}$ in Eq.~\eqref{tkecond}, with
Luttinger liquid parameter $g=\nu^{-1}$, the finite-$T$ corrections
at $T\ll T_{K}$ for the linear conductance tensor can be inferred
from Eq.~\eqref{tkecond} as well, 
\begin{equation}
G_{jk}=\frac{dI_{j}^{+}}{dV_{k}}=\nu\frac{e^{2}}{h}\left[1-(T/T_{K})^{2\Delta_{\mathrm{LIO}}-2}\right]\left(\frac{2}{M}-\delta_{jk}\right).
\end{equation}
Transport features are therefore basically identical to the Majorana-based
TKE, and also the PF-based strong-coupling point represents a local
quantum-critical point of non-Fermi-liquid type.

We close this section with two remarks. First, consider operators
of the form $e^{\pm i\sqrt{\nu}\left(\hat{\phi}_{j}(x>0,t)-\hat{\phi}_{j-1}(x'<0,t)\right)}$.
Such operators do not appear as perturbations within the Hamiltonian
in Eq.~\eqref{PFK}, which contains no direct tunneling processes
between different edges. In general, such couplings can appear and
destabilize the fixed point, even though these operators do not induce
transitions between different values of $\hat{C}_{jk}$, i.e., between
different ground-state minima of the potential in Eq.~\eqref{eq:H_PF_Kondo_cosines}.
Indeed, they couple the incoming and outgoing channels in the scattering
problem. Should the corresponding coupling strength be non-vanishing,
it will destabilize the fixed point below some energy scale. In practice,
these couplings (and the associated destabilization energy scale)
can be suppressed by arranging the respective edge parts far away
from each other.

Second, it is also instructive to consider operators $\hat{O}\sim e^{i\sqrt{\nu}\hat{\phi}_{m}(0)}\hat{\alpha}_{m}^{\dagger}$,
which commute with $\hat{C}{}_{jk}$ and have scaling dimension $\Delta=\nu/(2M)$,
with $\Delta<\nu/2$ for $M\geq2$. Therefore, if several couplings
$\lambda_{jk}$ in Eq.~\eqref{PFK} enter the strong-coupling regime
and approach a fixed point with $M'<M$ leads, the couplings to the
remaining leads will quickly catch up under the RG flow since they
are relevant. In fact, they are even more relevant than the original
couplings in Eq.~\eqref{PFK}.

\section{Conclusions}

\label{sec5}

In this work, we have proposed a parafermionic version of the topological
Kondo model previously suggested for a Majorana box \cite{Beri2012,Altland2013,Beri2013}.
Our generalization employs chiral fractional quantum Hall edge states
as leads, which are tunnel-coupled to parafermionic zero modes present
on a Coulomb-blockaded island, cf.~Ref.~\cite{Snizhko2018}. By
means of Abelian bosonization, a theoretical description of quantum
transport in such a multi-terminal quantum junction has been given
in both the weak- and the strong-coupling limit. In particular, we
have derived and discussed the one-loop RG equations. Our RG analysis
shows that the system flows towards an isotropic stable strong-coupling
fixed point. However, in contrast to the Majorana-based case, our
problem does not fall into the class of Kondo problems, see App.~\ref{appb}
for details.

The strong-coupling limit has then been analyzed by means of the approach
of Ref.~\cite{Ganeshan2016}, which yields controlled results within
the Abelian bosonization approach. It is remarkable that the resulting
conductance tensor is basically identical to the one of the Majorana-based
topological Kondo model, see Eq.~\eqref{tkecond}, even though no
continuous symmetries (and hence no Kondo screening processes) are
manifestly involved in the PF variant. Let us emphasize two particularly
noteworthy features of our $T=0$ result in Eq.~\eqref{t0cond}.
First, the isotropic partitioning of injected quasiparticle currents
into all outgoing leads is \emph{noiseless}, in analogy to previous
studies for different but related physical systems \cite{Beri2012,Altland2013,Beri2013,Zazunov2014,Altland2012,Altland2015}.
Second, consider the case that a current $I_{{\rm in}}$ is injected
only via lead 1. This lead then also serves as \emph{current extractor},
since the outgoing current $I_{1}^{+}$ has opposite sign as compared
to the injected current, with the universal ratio $I_{1}^{+}/I_{{\rm in}}=-(M-2)/M$.
By determining the leading irrelevant operators around the strong-coupling
point, we have also obtained the temperature-dependent corrections
to the conductance tensor and established that the strong-coupling
point represents a local quantum-critical point of non-Fermi liquid
type. Given the recent experimental advances in the field \cite{Ronen2018,Wu2018},
we hope that these predictions can soon be put to an experimental
test.

\acknowledgments We thank N. Andrei and A.M. Tsvelik for discussions.
We acknowledge funding by the Deutsche Forschungsgemeinschaft (Bonn)
within the network CRC TR 183 (project C01) and Grant No.~RO 2247/8-1,
by the IMOS Israel-Russia program, by the ISF, and the Italia-Israel
project QUANTRA.


\appendix

\section{On vertex operators}

\label{appA}

We here provide technical details related to Secs.~\ref{sec2} and
\ref{sec3}. We start by noting that in terms of the chiral boson
fields in Eqs.~\eqref{eq:H_edge} and \eqref{phicomm}, the fractional
quasiparticle operator for the $j$th lead is given by $\hat{\psi}_{j}(x,t)=V_{\sqrt{\nu},j}(x,t)$,
with the vertex operator 
\begin{equation}
V_{p,j}(x,t)=\left(\frac{L}{2\pi}\right)^{-p^{2}/2}:e^{ip\hat{\phi}_{j}(x,t)}:,\label{vertexop}
\end{equation}
where $:\ :$ denotes normal ordering and $L$ the system size. Using
$\tau_{c}$ as short-time cutoff and a set of conventional boson operators
$\{\hat{a}_{n,j}\}$ with momentum $k_{n}=2\pi n/L$ ($n\in\mathbb{N}$),
$\hat{\phi}_{j}$ has the mode decomposition \cite{Gogolin1998} 
\begin{eqnarray}
\hat{\phi}_{j}(x,t) & = & \hat{\phi}_{0,j}+\hat{Q}_{j}^{\mathrm{edge}}\frac{2\pi\left(x-vt\right)}{L\sqrt{\nu}}+i\sum_{n=1}^{\infty}\sqrt{\frac{2\pi}{Lk_{n}}}\label{eq:phi_j_quantized}\\
 & \times & \left(\hat{a}_{n,j}e^{ik_{n}(x-vt)}-\hat{a}_{n,j}^{\dagger}e^{-ik_{n}(x-vt)}\right)e^{-\tau_{c}vk_{n}/2},\nonumber 
\end{eqnarray}
where 
\begin{equation}
\left[\hat{Q}_{j}^{\mathrm{edge}},\hat{\phi}_{0,m}\right]=i\sqrt{\nu}\delta_{jm},\quad\left[\hat{\phi}_{0,j},\hat{\phi}_{0,m}\right]=i\pi\sgn(m-j).
\end{equation}
The operator $V_{p,j}$ in Eq.~\eqref{vertexop} has scaling dimension
$\Delta_{p}=p^{2}/2$. The OPE contractions required for deriving
the RG equations in Sec.~\ref{sec3} follow from the relation (with
$x'\to x$ and $t'\to t$) 
\begin{eqnarray}
 &  & V_{p,j}(x,t)V_{q,j}(x',t')=\left(\frac{L}{2\pi}\right)^{-(p+q)^{2}/2}\nonumber \\
 &  & \times\left[i(vt-x-vt'+x'-iv\tau_{c})\right]^{pq}\ :e^{i(p\hat{\phi}_{j}(x,t)+q\hat{\phi}_{j}(x',t'))}:\nonumber \\
 &  & =\left[i(vt-x-vt'+x'-iv\tau_{c})\right]^{pq}\nonumber \\
 &  & \quad\times\ V_{p+q,j}(x,t)\left[1+{\cal O}(x-x')+{\cal O}(t-t')\right].\label{OPEeq}
\end{eqnarray}
Note that $pq=\Delta_{p+q}-\Delta_{p}-\Delta_{q}$.

Using Eq.~\eqref{OPEeq} for $p=-q=\sqrt{\nu}$, and a regularization
with positive infinitesimal $\varepsilon$, we can also verify that
the potential scattering terms in Eq.~\eqref{eq:H_cot} can be neglected.
To that end, we write 
\begin{eqnarray}
\hat{\psi}_{j}^ {}(0)\hat{\psi}_{j}^{\dagger}(0) & = & \frac{1}{4\varepsilon^{2}}\int_{-\varepsilon}^{\varepsilon}dx\int_{-\varepsilon}^{\varepsilon}dx'\ \hat{\psi}_{j}^ {}(x)\hat{\psi}_{j}^{\dagger}(x')\label{eq1}\\
 & = & c_{0}+c_{1}\partial_{x}\hat{\phi}_{j}(0)+{\cal O}(c_{2}),\quad c_{m}\sim\varepsilon^{m-\nu}.\nonumber 
\end{eqnarray}
Now $c_{0}$ contributes only an unimportant (albeit divergent) constant
while, for $\nu<1$, all $c_{m\ge1}\to0$ for $\varepsilon\to0$.
For $\nu=1$, $c_{1}$ remains finite for $\varepsilon\to0$, but
the corresponding term in Eq.~\eqref{eq1} can be absorbed by the
transformation $\hat{\phi}_{j}(x)\to\hat{\phi}_{j}(x)+\tilde{c}_{j}\sgn(x)$
with $\tilde{c}_{j}\sim c_{1}$. Since similar statements hold for
$\hat{\psi}_{j}^{\dagger}(0)\hat{\psi}_{j}^ {}(0)$, we conclude that
for all $\nu\leq1$, potential scattering is indeed negligible.

\section{The (absent) symmetry of the leads}

\label{appb}

\subsubsection{Lead symmetries}

Kondo problems are characterized by coupling a set of leads, forming
a representation of a continuous symmetry, with a quantum impurity
sharing the same continuous symmetry. We here show that for our PF
generalization of the TKE with $\nu<1$, the leads do \emph{not} possess
a non-trivial continuous symmetry. As a consequence, the corresponding
model does not define a Kondo problem.

First of all, we note that for any $\nu\leq1$ and any number $M$
of leads, we have a $[U(1)]^{M}$ symmetry related to charge conservation
in each separate lead generated by the charge density operator $\frac{\sqrt{\nu}}{2\pi}\partial_{x}\hat{\phi}_{j}(x)$.
This is not the symmetry we are interested in, as it is Abelian. All
irreducible representations of an Abelian symmetry are necessarily
one-dimensional, and, therefore, generators of such a symmetry cannot
cause transitions of quasiparticles between the leads and the accompanying
changes in the impurity state. Hence we are looking for non-Abelian
symmetries of the leads, which should conserve total electric charge
and the scaling dimensions of transformed operators. A fractional
quasiparticle operator can thus only be transformed into a linear
combination of such operators, 
\begin{equation}
\hat{\psi}_{j}(x)\rightarrow\hat{\tilde{\psi}}_{j}(x)=\sum_{k}A_{jk}(x)\hat{\psi}_{k}(x).
\end{equation}
The symmetry should preserve the quasiparticle permutation relations
\eqref{psicomm}, which implies the conditions 
\begin{eqnarray}
 &  & \left(1-e^{i\pi\nu(\sgn(j-k)+\sgn(n-m))}e^{i\pi\nu\sgn(x-x')(\delta_{mn}-\delta_{jk})}\right)\nonumber \\
 &  & \quad\times A_{jm}(x)A_{kn}(x')=0\label{eq:symmtery_restr}
\end{eqnarray}
for arbitrary $(j,k,m,n$) indices and arbitrary $x\neq x'$. For
a continuous symmetry, we focus on infinitesimal transformations,
$A_{jk}(x)=\delta_{jk}+a_{jk}(x)$ with $|a_{jk}(x)|\ll1$, where
Eq.~\eqref{eq:symmtery_restr} yields (to linear order in $a_{jk}$
and putting $k=n$ and $j\neq m$) 
\begin{multline}
\left(1-e^{i\pi\nu(\sgn(j-k)+\sgn(k-m))}e^{i\pi\nu\sgn(x-x')(\delta_{mk}-\delta_{jk})}\right)\\
\times a_{jm}(x)=0.
\end{multline}
Here $x'$ remains as free parameter. Since the equation has to be
satisfied both at $x'>x$ and $x'<x$, one concludes that either $\nu=1$
or $a_{jm}(x)=0$. For $\nu\ne1$, this implies diagonal $A_{jm}(x)$
with Abelian symmetry, not mixing different edges. The above reasoning
thus constitutes a proof that no continuous non-Abelian lead symmetry
exists for $\nu\neq1$.

\subsubsection{Remarks on conformal field theory}

\begin{table}[t]
\begin{centering}
\begin{tabular}{|c|c|}
\hline 
$\nu$  & $\Delta_{n}$\tabularnewline
\hline 
\hline 
$1$  & $0,\frac{1}{3},1,\frac{4}{3},\frac{7}{3},3,...$ \tabularnewline
\hline 
$1/3$  & $0,\frac{1}{9},\frac{1}{3},\frac{4}{9},\frac{7}{9},1,\frac{4}{3},\frac{13}{9},\frac{16}{9},\frac{19}{9},\frac{7}{3},\frac{25}{9},3,...$\tabularnewline
\hline 
$1/5$  & $0,\frac{1}{15},\frac{1}{5},\frac{4}{15},\frac{7}{15},\frac{3}{5},\frac{4}{5},\frac{13}{15},\frac{16}{15},\frac{19}{15},\frac{7}{5},\frac{5}{3},\frac{9}{5},\frac{28}{15},\frac{31}{15},\frac{12}{5},...$\tabularnewline
\hline 
\end{tabular}
\par\end{centering}
\caption{\label{tab:neutral_scaling_dimensions} The list of smallest neutral
scaling dimensions of primary operators for $M=3$ leads at $\nu=1,\:1/3,\:1/5$.}
\end{table}

It is instructive to study this issue also from the conformal field
theory (CFT) \cite{Francesco1997} point of view. The leads are described
by a CFT for massless chiral bosons. According to Noether's theorem,
a continuous symmetry implies the existence of a conserved current
$J(x)$ generating the symmetry. In a chiral CFT, such currents must
have scaling dimension $\Delta=1$, since the total charge associated
with the conserved current, $\int dxJ(x)$, should not renormalize
under scaling. Therefore, if a continuous lead symmetry is present,
it must be generated by fields of scaling dimension $\Delta=1$. We
now separate $H_{{\rm edge}}$ into a charged and a neutral sector.
The charged sector is defined by the free boson field $\hat{\phi}_{c}(x)=\frac{1}{\sqrt{M}}\sum_{j=1}^{M}\hat{\phi}_{j}(x),$
with $\left[\hat{\phi}_{c}(x),\hat{\phi}_{c}(x')\right]=i\pi\sgn(x-x').$
The total charge density, expressed through 
\begin{equation}
\hat{\rho}_{c}(x)=\frac{\sqrt{\nu}}{2\pi}\partial_{x}\sum_{j=1}^{M}\hat{\phi}_{j}(x)=\frac{\sqrt{\nu M}}{2\pi}\partial_{x}\hat{\phi}_{c}(x),\label{eq:total_charge_density}
\end{equation}
generates the $U(1)_{c}$ symmetry responsible for total charge conservation.
Writing $H_{\mathrm{edge}}=H_{c}+H_{n}$ with $H_{c}=\frac{v}{4\pi}\int_{-L/2}^{+L/2}dx\left(\partial_{x}\hat{\phi}_{c}(x)\right)^{2}$
and 
\begin{equation}
H_{n}=\frac{v}{4\pi}\sum_{j=1}^{M}\int_{-L/2}^{+L/2}dx\left(\partial_{x}\hat{\phi}_{j}(x)-\frac{1}{\sqrt{M}}\partial_{x}\hat{\phi}_{c}(x)\right)^{2},\label{eq:H_n}
\end{equation}
the degrees of freedom of the neutral sector do not carry electric
charge, $\left[\hat{\phi}_{j}(x)-\frac{1}{\sqrt{M}}\hat{\phi}_{c}(x),\hat{\rho}_{c}(y)\right]=0.$
Each operator in the theory can then be decomposed into a product
of charged and neutral parts, 
\begin{equation}
\hat{O}(x)=e^{iq\hat{\phi}_{c}(x)/\sqrt{\nu M}}\otimes\hat{O}_{n}(x),
\end{equation}
with scaling dimension $\Delta_{\hat{O}}=\Delta_{c}+\Delta_{n},$
where $\Delta_{c}=q^{2}/(2\nu M)$. The spectrum of neutral scaling
dimensions $\Delta_{n}$ thus follows by computing $\Delta_{\hat{O}}-\Delta_{c}$
for all primary operators. Operators with $\Delta_{n}=1$ and $q=0$
are candidates for generators of hidden symmetries in the neutral
sector. Apart from $\partial_{x}\hat{\phi}_{j}(x)-\frac{1}{\sqrt{M}}\partial_{x}\hat{\phi}_{c}(x)$,
primary fields are given by 
\begin{equation}
\hat{O}_{\{n_{j}\}}(x)=e^{i\sum_{j=1}^{M}n_{j}\hat{\phi}_{j}(x)\sqrt{\nu}},\label{eq:bulk_operators}
\end{equation}
with charge $q=\nu\sum_{j}n_{j}$ and scaling dimension $\Delta=\nu\sum_{j}n_{j}^{2}/2$,
where all $n_{j}\in\mathbb{Z}$. The spectrum of neutral scaling dimensions
follows as 
\begin{eqnarray}
\Delta_{n} & = & \frac{\nu}{2}\left(\sum_{j}n_{j}^{2}-\frac{1}{M}\left[\sum_{j}n_{j}\right]^{2}\right)\nonumber \\
 & = & \frac{\nu}{2}\sum_{j}\left(n_{j}-\frac{\sum_{k=1}^{M}n_{k}}{M}\right)^{2}\geq0.\label{eq:neut_scal_dim_bulk}
\end{eqnarray}
For simplicity, we focus from now on the case of $M=3$ leads. The
smallest scaling dimensions in the neutral sector, calculated from
Eq.~\eqref{eq:neut_scal_dim_bulk}, are listed for $\nu=1$, $1/3$,
and $1/5$ in Table~\ref{tab:neutral_scaling_dimensions}. First,
note that for $\nu=1/5$, there are no operators with $\Delta_{n}=1$,
and, therefore, no continuous symmetries exist in the neutral sector
apart from the $[U(1)]^{2}$ symmetry generated by $\partial_{x}\hat{\phi}_{j}(x)-\frac{1}{\sqrt{M}}\partial_{x}\hat{\phi}_{c}(x)$.

The case of $\nu=1$ is equivalent to the TKE, where one expects an
so$(3)=$~su(2) symmetry \cite{Beri2012}. Indeed, all operators
with scaling dimension $\Delta=1$ are expressed as $J_{jk}(x)=\hat{\psi}_{j}^{\dagger}(x)\hat{\psi}_{k}^ {}(x)$
in terms of the electron operators $\hat{\psi}_{j}(x)$. The total
charge density in Eq.~\eqref{eq:total_charge_density} corresponds
to $\hat{\rho}_{c}(x)\sim\sum_{j}\hat{\psi}_{j}^{\dagger}(x)\hat{\psi}_{j}^ {}(x)$,
while the remaining eight currents generate a symmetry of the neutral
sector. These eight currents obey the su$(3)_{1}$ Kac-Moody (KM)
algebra, and the theory of three $\nu=1$ leads can be described as
su$(3)_{1}$ Wess-Zumino-Witten (WZW) CFT. A subalgebra of this algebra
constitutes the su$(2)_{4}$ KM algebra, and the leads can also be
described by the su$(2)_{4}$ WZW model, which ultimately provides
a description for the strong-coupling fixed point of the TKE \cite{Beri2012}.

The most interesting case for us is $\nu=1/3$. Apart from $\hat{\rho}_{c}(x)$,
there are again eight operators with $\Delta=1$, which obey the same
commutation relations as the su$(3)$-generating Gell-Mann matrices, 
\begin{widetext}
\begin{equation}
\begin{array}{cc}
\lambda^{1}=\begin{pmatrix}0 & 1 & 0\\
1 & 0 & 0\\
0 & 0 & 0
\end{pmatrix}\leftrightarrow J^{1}(x)=J_{3}^{+}(x)+J_{3}^{-}(x), & \lambda^{2}=\begin{pmatrix}0 & -i & 0\\
i & 0 & 0\\
0 & 0 & 0
\end{pmatrix}\leftrightarrow J^{2}(x)=-iJ_{3}^{+}(x)+iJ_{3}^{-}(x),\\
\lambda^{4}=\begin{pmatrix}0 & 0 & 1\\
0 & 0 & 0\\
1 & 0 & 0
\end{pmatrix}\leftrightarrow J^{4}(x)=J_{2}^{-}(x)+J_{2}^{+}(x), & \lambda^{5}=\begin{pmatrix}0 & 0 & -i\\
0 & 0 & 0\\
i & 0 & 0
\end{pmatrix}\leftrightarrow J^{5}(x)=-iJ_{2}^{-}(x)+iJ_{2}^{+}(x),\\
\lambda^{6}=\begin{pmatrix}0 & 0 & 0\\
0 & 0 & 1\\
0 & 1 & 0
\end{pmatrix}\leftrightarrow J^{6}(x)=J_{1}^{+}(x)+J_{1}^{-}(x), & \lambda^{7}=\begin{pmatrix}0 & 0 & 0\\
0 & 0 & -i\\
0 & i & 0
\end{pmatrix}\leftrightarrow J^{7}(x)=-iJ_{1}^{+}(x)+iJ_{1}^{-}(x),\\
\lambda^{3}=\begin{pmatrix}1 & 0 & 0\\
0 & -1 & 0\\
0 & 0 & 0
\end{pmatrix}\leftrightarrow J^{3}(x)=\frac{\partial(\hat{\phi}_{1}+\hat{\phi}_{2}-2\hat{\phi}_{3})(x)}{\sqrt{3}}, & \lambda^{8}=\frac{1}{\sqrt{3}}\begin{pmatrix}1 & 0 & 0\\
0 & 1 & 0\\
0 & 0 & -2
\end{pmatrix}\leftrightarrow J^{8}(x)=\partial_{x}(\hat{\phi}_{2}-\hat{\phi}_{1})(x),
\end{array}\label{eq:Laughlin_su(3)_currents}
\end{equation}
\end{widetext}

where 
\begin{equation}
J_{j}^{\pm}(x)=l^{-1}e^{\pm i\left(3\hat{\phi}_{j}-\hat{\phi}_{1}-\hat{\phi}_{2}-\hat{\phi}_{3}\right)/\sqrt{3}},
\end{equation}
with cutoff length $l$. Note the apparent strangeness in definitions
of $J^{3}$, $J^{8}$, and $J^{5}$, as compared to $J^{2}$ and $J^{7}$.
Moreover, these currents obey the su$(3)_{1}$ KM algebra. $H_{n}$
in Eq.~\eqref{eq:H_n} can be expressed in terms of these currents,
and then coincides with the Hamiltonian of the su$(3)_{1}$ WZW model
\cite{Francesco1997}. One would therefore expect that the leads have
su(3) symmetry and are described by the WZW model. However, this is
not the case. One way to see this is to note that the scaling dimensions
$\Delta_{n}=\frac{1}{9},\frac{4}{9},\frac{7}{9}$ in~Table~\ref{tab:neutral_scaling_dimensions}
are absent in the spectrum of the su$(3)_{1}$ WZW model. Moreover,
the currents $J^{k}(x)$ do not act as a symmetry on the operators
in the theory.

Indeed, consider the OPE of a current with a quasiparticle operator
$\hat{\psi}_{j}(x)\sim e^{i\hat{\phi}_{j}(x)/\sqrt{3}}$. For example,
$J_{1}^{+}(x)\times e^{i\hat{\phi}_{2}(y)/\sqrt{3}}\sim(x-y)^{-1/3}e^{i\left(2\hat{\phi}_{1}-\hat{\phi}_{3}\right)(y)/\sqrt{3}}$,
thus mapping an operator with $\Delta_{n}=\frac{1}{9}$ onto one with
$\Delta_{n}=\frac{7}{9}$. Operators with `correct' scaling dimensions
in the su$(3)_{1}$ WZW model are mapped correctly. For instance,
operators of scaling dimension $\Delta_{n}=\frac{1}{3}$, which are
given by $\hat{\psi}_{j}^{\dagger}(x)\hat{\psi}_{k}(x)$ with $j\neq k$,
are mapped in agreement with the fundamental $\left(\hat{\psi}_{1}^{\dagger}(x)\hat{\psi}_{2}(x),\hat{\psi}_{2}^{\dagger}(x)\hat{\psi}_{3}(x),\hat{\psi}_{3}^{\dagger}(x)\hat{\psi}_{1}(x)\right)$
and the anti-fundamental (the other three) representations of su(3).
However, the currents and the operators do not commute at distant
points, e.g., 
\begin{equation}
J_{3}^{+}(x)\hat{\psi}_{3}^{\dagger}\hat{\psi}_{1}^{\thinspace}(y)=\hat{\psi}_{3}^{\dagger}\hat{\psi}_{1}^{\thinspace}(y)J_{3}^{+}(x)e^{i\pi(\sgn(x-y)-1/3)}.
\end{equation}
This last statement means that operators $\hat{\psi}_{j}^{\dagger}(x)\hat{\psi}_{k}(x)$
are not local with respect to currents $J^{k}(x)$. On the other hand,
all the operators in the theory of the leads are local with respect
to electron operators $e^{i\sqrt{3}\hat{\phi}_{j}(x)}$ (by construction
of the FQH edges). The two models, the theory of the leads and su$(3)_{1}$
WZW model, are therefore `almost' the same: they have the same central
charge and even the same Hamiltonian, yet they have a different spectrum.
The origin of this difference appears to come from different locality
notions. This is not a unique situation: the same relation is present
between the theory of free Majorana fermions in 1+1 dimensions and
the $\mathcal{M}(4,3)$ minimal CFT model describing the critical
point of the two-dimensional (2D) Ising model \cite{Francesco1997}.
Both models have central charge $1/2$, yet the spin operator $\sigma$
of scaling dimension $\Delta=1/16$ is non-local with respect to the
Majorana fermion and thus absent from the former theory. The relation
between the models is evident through Onsager's solution of the 2D
Ising model. Further studies of such `locality-distinguished' CFTs
may uncover similar relations and possibly allow for full solutions
of models whose critical point is described by the su$(3)_{1}$ WZW
model. We expect that for $M>3$, similar `almost realized' symmetries
will be encountered for $\nu=1/3$ and for $\nu=1/5$.

\section{On the strong-coupling solution}

\label{appc}

We here provide technical details about our strong-coupling solution
in Sec.~\ref{sec4}. In particular, we derive the expressions for
the field operators quoted in Eqs.~\eqref{phiii} and \eqref{phii}.
Using the approach of Ref.~\cite{Ganeshan2016}, the low-energy Hamiltonian
is given by 
\begin{equation}
H_{\mathrm{LE}}=H_{0}-\frac{1}{2}\sum_{j,k=2}^{M}\mathcal{N}_{jk}\hat{\Pi}_{j}\hat{\Pi}_{k}.\label{hle}
\end{equation}
Here we use the integer-valued operators $\hat{D}_{j}\equiv\hat{C}_{1j}$
(with $j=2,\ldots,M$), and the conjugate operators 
\begin{equation}
\hat{\Pi}_{j}=\frac{1}{2\pi i}\sum_{k=2}^{M}\mathcal{M}_{jk}\left[\hat{D}_{k},H_{0}\right],\label{Pidef}
\end{equation}
with symmetric matrices $\mathcal{N}$ and $\mathcal{M}$ given by
\begin{equation}
\mathcal{N}_{jk}=-\frac{1}{4\pi^{2}}\left[\hat{D}_{k},\left[\hat{D}_{j},H_{0}\right]\right],\quad\mathcal{M}=\mathcal{N}^{-1}.\label{nmdef}
\end{equation}
Noting that $\left[\hat{D}_{j},\hat{\Pi}_{k}\right]=2\pi i\delta_{jk}$,
the operator $e^{\pm i\hat{\Pi}_{j}}$ effectively shifts $\hat{D}_{j}\to\hat{D}_{j}\pm1$.

In order to implement the approach of Ref.~\cite{Ganeshan2016},
we discretize the spatial coordinate $x=z\varepsilon$ in units of
a small spacing $\varepsilon$ (with $z\in\mathbb{Z}$), where 
\begin{equation}
H_{0}=\sum_{j=1}^{M}\frac{v}{4\pi\varepsilon}\sum_{z}\left(\hat{\phi}_{j}(z\varepsilon+\varepsilon)-\hat{\phi}_{j}(z\varepsilon)\right)^{2}.\label{eq:H0_regularized}
\end{equation}
Using the commutation relations \eqref{phicomm}, the matrices in
Eq.~\eqref{nmdef} take the form 
\begin{equation}
\mathcal{N}_{jk}=\frac{v\nu}{4\pi\varepsilon}\left(1+\delta_{jk}\right),\quad\mathcal{M}_{jk}=\frac{4\pi\varepsilon}{v\nu}\left(\delta_{jk}-\frac{1}{M}\right),
\end{equation}
while Eq.~\eqref{Pidef} yields 
\begin{equation}
\hat{\Pi}_{j}=\frac{1}{\sqrt{\nu}}\left(\hat{\phi}_{j}(-\varepsilon)-\hat{\phi}_{j}(\varepsilon)-\sum_{k=1}^{M}\frac{\hat{\phi}_{k}(-\varepsilon)-\hat{\phi}_{k}(\varepsilon)}{M}\right).
\end{equation}
The low-energy Hamiltonian \eqref{hle} is thus given by 
\begin{eqnarray}
H_{\mathrm{LE}} & = & H_{0}-\frac{v}{8\pi\varepsilon}\sum_{k=1}^{M}\left(\hat{\phi}_{k}(\varepsilon)-\hat{\phi}_{k}(-\varepsilon)\right)^{2}\nonumber \\
 & + & \frac{v}{8\pi\varepsilon M}\left[\sum_{k=1}^{M}\left(\hat{\phi}_{k}(\varepsilon)-\hat{\phi}_{k}(-\varepsilon)\right)\right]^{2}.\label{eq:H_LE}
\end{eqnarray}
Since $H_{{\rm LE}}$ is quadratic in the boson fields, it can easily
be diagonalized. To that end, consider the equations of motion. First,
for $|z|\ge2$, 
\begin{equation}
\partial_{t}\hat{\phi}_{j}(z\varepsilon)=-v\frac{\hat{\phi}_{k}(z\varepsilon+\varepsilon)-\hat{\phi}_{k}(z\varepsilon-\varepsilon)}{2\varepsilon}.\label{eq:EOM_x}
\end{equation}
For $z=\pm1$, one gets 
\begin{eqnarray}
\partial_{t}\hat{\phi}_{j}(\pm\varepsilon) & = & \mp v\frac{\hat{\phi}_{j}(\pm2\varepsilon)-\hat{\phi}_{j}(0)}{2\varepsilon}+\frac{v}{2}\Biggl(\frac{\hat{\phi}_{j}(\varepsilon)-\hat{\phi}_{j}(-\varepsilon)}{2\varepsilon}\nonumber \\
 & - & \frac{1}{M}\sum_{k=1}^{M}\frac{\hat{\phi}_{k}(\varepsilon)-\hat{\phi}_{k}(-\varepsilon)}{2\varepsilon}\Biggr),\label{eq:EOM_eps}
\end{eqnarray}
while for $z=0$, we have 
\begin{equation}
\partial_{t}\hat{\phi}_{j}(0)=-\frac{v}{M}\sum_{k=1}^{M}\frac{\hat{\phi}_{k}(\varepsilon)-\hat{\phi}_{k}(-\varepsilon)}{2\varepsilon}.\label{eq:EOM_0}
\end{equation}
In addition, the constraints \eqref{constraint} have to be satisfied
at all times, where $\hat{\beta}_{jk}$ and $\hat{C}_{jk}$ do not
depend on time since each of them commutes with $H_{\mathrm{LE}}$.
Taking the limit $\varepsilon\to0$, Eq.~\eqref{eq:EOM_x} implies
the dispersion relation $\omega=vq$. At given frequency $\omega$,
we then obtain the time-dependent solutions 
\begin{equation}
\hat{\phi}_{j}(z\varepsilon,t)=\begin{cases}
u_{j}^{-}(\omega)e^{iq(z\varepsilon+\varepsilon)-i\omega t} & ,\:z\leq-1,\\
u_{j}^{+}(\omega)e^{iq(z\varepsilon-\varepsilon)-i\omega t} & ,\:z\geq1,\\
u_{j}^{0}(\omega)e^{-i\omega t} & ,\:z=0.
\end{cases}\label{finalfield}
\end{equation}
For $\varepsilon\to0$, Eq.~\eqref{eq:EOM_0} together with Eq.~\eqref{constraint}
yields 
\begin{equation}
u_{j}^{0}(\omega)=u_{k}^{0}(\omega)+\delta_{\omega,0}\pi\sqrt{\nu^{-1}}\left(\hat{\beta}_{jk}-2\hat{C}_{jk}\right),\label{eq:bc_0}
\end{equation}
and by combining Eqs.~\eqref{eq:EOM_eps} and \eqref{eq:EOM_0},
we obtain 
\begin{equation}
2u_{j}^{0}(\omega)=u_{j}^{+}(\omega)+u_{j}^{-}(\omega),\quad\sum_{j=1}^{M}\left(u_{j}^{+}(\omega)-u_{j}^{-}(\omega)\right)=0.\label{eq:bc}
\end{equation}
Using also Eq.~\eqref{eq:bc}, we finally arrive at 
\begin{eqnarray}
u_{j}^{+}(\omega) & = & -u_{j}^{-}(\omega)+\frac{2}{M}\sum_{k=1}^{M}u_{k}^{-}(\omega)+\\
 & + & \delta_{\omega,0}\frac{2}{M}\pi\sqrt{\nu^{-1}}\sum_{k=1,k\neq j}^{M}\left(\hat{\beta}_{jk}-2\hat{C}_{jk}\right),\nonumber 
\end{eqnarray}
where $u_{j}^{0}(\omega)=[u_{j}^{+}(\omega)+u_{j}^{-}(\omega)]/2$.
These relations directly yield Eqs.~\eqref{phiii} and \eqref{phii}.


\begin{thebibliography}{10}
\bibitem{Alicea2012} J. Alicea, Rep. Prog. Phys. \textbf{75}, 076501
(2012). 

\bibitem{Leijnse2012} M. Leijnse and K. Flensberg, Semicond. Sci.
Techn. \textbf{27}, 124003 (2012). 

\bibitem{Beenakker2013} C.W.J. Beenakker, Annu. Rev. Condens. Matt.
Phys. \textbf{4}, 113 (2013). 

\bibitem{Lutchyn2018} R.M. Lutchyn, E.P.A.M. Bakkers, L.P. Kouwenhoven,
P. Krogstrup, C.M. Marcus, and Y. Oreg, arXix:1707.04899. 

\bibitem{Nayak2008} C. Nayak, S.H. Simon, A. Stern, M. Freedman,
and S. Das Sarma, Rev. Mod. Phys. \textbf{80}, 1083 (2008). 

\bibitem{Alicea2016} J. Alicea and P. Fendley, Annu. Rev. Condens.
Matter Phys. \textbf{7}, 119 (2016). 

\bibitem{Sarma2015} S. Das Sarma, M. Freedman, and C. Nayak, npj
Quantum Information \textbf{1}, 15001 (2015). 

\bibitem{Mourik2012} V. Mourik, K. Zuo, S.M. Frolov, S.R. Plissard,
E.P.A. Bakkers, and L.P. Kouwenhoven, Science \textbf{336}, 1003 (2012). 

\bibitem{Deng2012} M.T. Deng, C.L. Yu, G.Y. Huang, M. Larsson, P.
Caroff, and H.Q. Xu, Nano Lett. \textbf{12}, 6414 (2012). 

\bibitem{Das2012} A. Das, Y. Ronen, Y. Most, Y. Oreg, M. Heiblum,
and H. Shtrikman, Nature Phys. \textbf{8}, 887 (2012). 

\bibitem{Rokhinson2012} L.P. Rokhinson, X. Liu, and J.K. Furdyna,
Nature Phys. \textbf{8}, 795 (2012). 

\bibitem{Yazdani2014} S. Nadj-Perge, I.K. Drozdov, J. Li, H. Chen,
S. Jeon, J. Seo, A.H. MacDonald, B.A. Bernevig, and A. Yazdani, Science
\textbf{346}, 602 (2014). 

\bibitem{Franke2015} M. Ruby, F. Pientka, Y. Peng, F. von Oppen,
B.W. Heinrich, and K.J. Franke, Phys. Rev. Lett. \textbf{115}, 197204
(2015). 

\bibitem{Sun2016} H.H. Sun, K.W. Zhang, L.H. Hu, C. Li, G.Y. Wang,
H.Y. Ma, Z.A. Xu, C.L. Gao, D.D. Guan, Y.Y. Li, C. Liu, D. Qian, Y.
Zhou, L. Fu, S.C. Li, F.C. Zhang, and J.F. Jia, Phys. Rev. Lett. \textbf{116},
257003 (2016). 

\bibitem{Albrecht2016} S.M. Albrecht, A.P. Higginbotham, M. Madsen,
F. Kuemmeth, T.S. Jespersen, J. Nyg{å}rd, P. Krogstrup, and C.M.
Marcus, Nature \textbf{531}, 206 (2016). 

\bibitem{Deng2016} M.T. Deng, S. Vaitiekenas, E.B. Hansen, J. Danon,
M. Leijnse, K. Flensberg, J. Nyg{å}rd, P. Krogstrup, and C.M. Marcus,
Science \textbf{354}, 1557 (2016). 

\bibitem{Guel2017} Ö. Gül, H. Zhang, F.K. de Vries, J. van Veen,
K. Zuo, V. Mourik, S. Conesa-Boj, M.P. Nowak, D.J. van Woerkom, M.
Quintero-P{é}rez, M.C. Cassidy, A. Geresdi, S. Koelling, D. Car,
S.R. Plissard, E.P.A.M. Bakkers, and L.P. Kouwenhoven, Nano Lett.
\textbf{17}, 2690 (2017). 

\bibitem{Zhang2017} H. Zhang, {Ö}. G{ü}l, S. Conesa-Boj, M. Nowak,
M. Wimmer, K. Zuo, V. Mourik, F.K. de Vries, J. van Veen, M.W.A. de
Moor, J.D.S. Bommer, D.J. van Woerkom, D. Car, S.R. Plissard, E.P.A.M.
Bakkers, M. Quintero-Perez, M.C. Cassidy, S. Koelling, S. Goswami,
K. Watanabe, T. Taniguchi, and L.P. Kouwenhoven, Nature Commun. \textbf{8},
16025 (2017). 

\bibitem{Albrecht2017} S.M. Albrecht, E.B. Hansen, A.P. Higginbotham,
F. Kuemmeth, T.S. Jespersen, J. Nyg{å}rd, P. Krogstrup, J. Danon,
K. Flensberg and C.M. Marcus, Phys. Rev. Lett. \textbf{118}, 137701
(2017). 

\bibitem{Nichele2017} F. Nichele, A.C.C. Drachmann, A.M. Whiticar,
E.C.T. O'Farrell, H.J. Suominen, A. Fornieri, T. Wang, G.C. Gardner,
C. Thomas, A.T. Hatke, P. Krogstrup, M.J. Manfra, K. Flensberg, and
C.M. Marcus, Phys. Rev. Lett. \textbf{119}, 136803 (2017). 

\bibitem{Suominen2017} H.J. Suominen, M. Kjaergaard, A.R. Hamilton,
J. Shabani, C.J. Palmstr{ø}m, C.M. Marcus, and F. Nichele, Phys.
Rev. Lett. \textbf{119}, 176805 (2017). 

\bibitem{Gazi2017} S. Gazibegovich, D. Car, H. Zhang, S.C. Balk,
J.A. Logan, M.W.A. de Moor, M.C. Cassidy, R. Schmits, D. Xu, G. Wang,
P. Krogstrup, R.L.M. Op het Veld, J. Shen, D. Bouman, B. Shojaei,
D. Pennachio, J.S. Lee, P.J. van Veldhoven, S. Koelling, M.A. Verheijen,
L.P. Kouwenhoven, C.J. Palmstr{ø}m, and E.P.A.M. Bakkers, Nature
\textbf{548}, 434 (2017). 

\bibitem{Feldman2017} B.E. Feldman, M.T. Randeria, J. Li, S. Jeon,
Y. Xie, Z. Wang, I.K. Drozdov, B. Andrei Bernevig, and A. Yazdani,
Nat. Phys. \textbf{13}, 286 (2017). 

\bibitem{Deacon2017} R.S. Deacon, J. Wiedenmann, E. Bocquillon, F.
Dom{í}nguez, T.M. Klapwijk, P. Leubner, C. Br{ü}ne, E.M. Hankiewicz,
S. Tarucha, K. Ishibashi, H. Buhmann, and L.W. Molenkamp, Phys. Rev.
X \textbf{7}, 021011 (2017). 

\bibitem{Ronen2018} Y. Ronen, Y. Cohen, D. Banditt, M. Heiblum, and
V. Umansky, Nature Phys. \textbf{14}, 411 (2018). 

\bibitem{Wu2018} T. Wu, Z. Wan, A. Kazakov, Y. Wang, G. Simion, J.
Liang, K.W. West, K. Baldwin, L.N. Pfeiffer, Y. Lyanda-Geller, L.P.
Rokhinson, Phys. Rev. B \textbf{97}, 245304 (2018). 

\bibitem{Lindner2012} N.H. Lindner, E. Berg, G. Refael, and A. Stern,
Phys. Rev. X \textbf{2}, 041002 (2012). 

\bibitem{Cheng2012} M. Cheng, Phys. Rev. B \textbf{86}, 195126 (2012). 

\bibitem{Clarke2013} D.J. Clarke, J. Alicea, and K. Shtengel, Nature
Commun. \textbf{4}, 1348 (2013). 

\bibitem{Burrello2013} M. Burrello, B. van Heck, and E. Cobanera,
Phys. Rev. B \textbf{87}, 195422 (2013). 

\bibitem{Vaezi2013} A. Vaezi, Phys. Rev. B \textbf{87}, 035132 (2013). 

\bibitem{Zhang2014} F. Zhang and C.L. Kane, Phys. Rev. Lett. \textbf{113},
036401 (2014). 

\bibitem{Mong2014} R.S.K. Mong, D.J. Clarke, J. Alicea, N.H. Lindner,
P. Fendley, C. Nayak, Y. Oreg, A. Stern, E. Berg, K. Shtengel, and
M.P.A. Fisher, Phys. Rev. X \textbf{4}, 011036 (2014). 

\bibitem{Clarke2014} D.J. Clarke, J. Alicea, and K. Shtengel, Nature
Phys. \textbf{10}, 877 (2014). 

\bibitem{Barkeshli2014a} M. Barkeshli, Y. Oreg, and X.L. Qi, arXiv:1401.3750. 

\bibitem{Barkeshli2014b} M. Barkeshli and X.L. Qi, Phys. Rev. X \textbf{4},
041035 (2014). 

\bibitem{Klinovaja2014} J. Klinovaja and D. Loss, Phys. Rev. Lett.
\textbf{112}, 246403 (2014). 

\bibitem{Klinovaja2014b} J. Klinovaja, A. Yacoby, and D. Loss, Phys.
Rev. B \textbf{90}, 155447 (2014). 

\bibitem{Cheng2015} M. Cheng and R.M. Lutchyn, Phys. Rev. B \textbf{92},
134516 (2015). 

\bibitem{Alicea2015a} J. Alicea and A. Stern, Phys. Scr. \textbf{T164},
14006 (2015). 

\bibitem{Kim2017} Y. Kim, D.J. Clarke, and R.M. Lutchyn, Phys. Rev.
B \textbf{96}, 041123 (2017). 

\bibitem{Snizhko2018} K. Snizhko, R. Egger, and Y. Gefen, Phys. Rev.
B \textbf{97}, 081405(R) (2018). 

\bibitem{Pachos2018} K. Meichanetzidis, C.J. Turner, A. Farjami,
Z. Papi{\'{c}}, and J.K. Pachos, Phys. Rev. B \textbf{97}, 125104
(2018). 

\bibitem{Chew2018} A. Chew, D. Mross, and J. Alicea, arXiv:1802.04809. 

\bibitem{Lee2017} G. Lee, K. Huang, D.K. Efetov, D.S. Wei, S. Hart,
T. Taniguchi, K. Watanabe, A. Yacoby, and P. Kim, Nature Phys. \textbf{13},
693 (2017). 

\bibitem{Lee2017a} J.S. Lee, B. Shojaei, M. Pendharkar, A.P. McFadden,
Y. Kim, H.J. Suominen, M. Kjaergaard, F. Nichele, C.M. Marcus, and
C.J. Palmstr{ø}m, arXiv:1705.05049. 

\bibitem{Fu2010} L. Fu, Phys. Rev. Lett. \textbf{104}, 056402 (2010). 

\bibitem{Zazunov2011} A. Zazunov, A.L. Yeyati, and R. Egger, Phys.
Rev. B \textbf{84}, 165440 (2011). 

\bibitem{Hutzen2012} R. H{ü}tzen, A. Zazunov, B. Braunecker, A.L.
Yeyati, and R. Egger, Phys. Rev. Lett. \textbf{109}, 166403 (2012). 

\bibitem{Aasen2016} D. Aasen, M. Hell, R.V. Mishmash, A.Higginbotham,
J. Danon, M. Leijnse, T.S. Jespersen, J.A. Folk, C.M. Marcus, K. Flensberg,
and J. Alicea, Phys. Rev. X \textbf{6}, 031016 (2016). 

\bibitem{Plugge2016} S. Plugge, L.A. Landau, E. Sela, A. Altland,
K. Flensberg, and R. Egger, Phys. Rev. B \textbf{94}, 174514 (2016). 

\bibitem{Landau2016} L.A. Landau, S. Plugge, E. Sela, A. Altland,
S.M. Albrecht, and R. Egger, Phys. Rev. Lett. \textbf{116}, 050501
(2016). 

\bibitem{Plugge2017} S. Plugge, A. Rasmussen, R. Egger, and K. Flensberg,
New J. Phys. \textbf{19}, 012001 (2017). 

\bibitem{Karzig2017} T. Karzig, C. Knapp, R.M. Lutchyn, P. Bonderson,
M.B. Hastings, C. Nayak, J. Alicea, K. Flensberg, S. Plugge, Y. Oreg,
C.M. Marcus, and M.H. Freedman, Phys. Rev. B \textbf{95}, 235305 (2017). 

\bibitem{Litinski2017} D. Litinski, M.S. Kesselring, J. Eisert, and
F. von Oppen, Phys. Rev. X \textbf{7}, 031048 (2017). 

\bibitem{Beri2012} B. B{é}ri and N.R. Cooper, Phys. Rev. Lett.
\textbf{109}, 156803 (2012). 

\bibitem{Altland2013} A. Altland and R. Egger, Phys. Rev. Lett. \textbf{110},
196401 (2013). 

\bibitem{Beri2013} B. B{é}ri, Phys. Rev. Lett. \textbf{110}, 216803
(2013). 

\bibitem{Crampe2013} N. Cramp{é} and A. Trombettoni, Nucl. Phys.
\textbf{B871}, 526 (2013). 

\bibitem{Tsvelik2013} A.M. Tsvelik, Phys. Rev. Lett. \textbf{110},
147202 (2013). 

\bibitem{Zazunov2014} A. Zazunov, A. Altland, and R. Egger, New J.
Phys. \textbf{16}, 015010 (2014). 

\bibitem{Altland2014} A. Altland, B. B{é}ri, R. Egger, and A.M.
Tsvelik, Phys. Rev. Lett. \textbf{113}, 076401 (2014). 

\bibitem{Galpin2014} M.R. Galpin, A.K. Mitchell, J. Temaismithi,
D.E. Logan, B. B{é}ri, and N.R. Cooper, Phys. Rev. B \textbf{89},
045143 (2014). 

\bibitem{Buccheri2015} F. Buccheri, H. Babujian, V. E. Korepin, P.
Sodano, and A. Trombettoni, Nucl. Phys. \textbf{B896}, 52 (2015). 

\bibitem{Zazunov2017} A. Zazunov, F. Buccheri, P. Sodano, and R.
Egger, Phys. Rev. Lett. \textbf{118}, 057001 (2017). 

\bibitem{Wen1991} X.G. Wen, Phys. Rev. B \textbf{44}, 5708 (1991). 

\bibitem{Kane1992a} C.L. Kane and M.P.A. Fisher, Phys. Rev. Lett.
\textbf{68}, 1220 (1992). 

\bibitem{Gogolin1998} A.O. Gogolin, A.A. Nersesyan, and A.M. Tsvelik,
\textit{Bosonization and Strongly Correlated Systems} (Cambridge University
Press, Cambridge UK, 1998). 

\bibitem{Altland2010} A. Altland and B. Simons, \textit{Condensed
Matter Field Theory}, 2nd ed. (Cambridge University Press, Cambridge
UK, 2010). 

\bibitem{footnote1} For comparison with the TKE (i.e., $n=2$), let
us note that the algebra su$(d)$, with $d=2^{[(M-1)/2]}$, contains
the expected subalgebra so$(M)$ for $M\ge7$. In the TKE, Majorana
bilinears generate this subalgebra, and leads are only coupled to
this subalgebra \cite{Beri2012}. For $M<7$, on the contrary, the
Hilbert space dimension is $d<M$, meaning that the system forms a
representation of so$(M)$ which is smaller than the fundamental one.
The quoted value for $d$ follows by noting that one needs at least
$[(M+1)/2]$ PF pairs on the box to couple to $M$ external edges
(leads), where constraining the total $\mathbb{Z}_{n}$ charge of
the box is equivalent to removing one PF pair, cf.~Ref.~\cite{Snizhko2018}. 

\bibitem{Dong2017} Z.Y. Dong, S.L. Yu, and J.X. Li, Phys. Rev. B
\textbf{96}, 245114 (2017). 

\bibitem{Goldman1995} V.J. Goldman and B. Su, Science \textbf{267},
1010 (1995). 

\bibitem{Goldman1996} V.J. Goldman, Surf. Sci. \textbf{361-362},
1 (1996). 

\bibitem{Picciotto1997} R. de Picciotto, M. Reznikov, M. Heiblum,
V. Umansky, G. Bunin, and D. Mahalu, Nature \textbf{389}, 162 (1997). 

\bibitem{Saminadayar1997} L. Saminadayar, D.C. Glattli, Y. Jin, and
B. Etienne, Phys. Rev. Lett. \textbf{79}, 2526 (1997). 

\bibitem{Goldman1997} V.J. Goldman, Physica E \textbf{1}, 15 (1997). 

\bibitem{Maasilta1997} I.J. Maasilta and V.J. Goldman, Phys. Rev.
B \textbf{55}, 4081 (1997). 

\bibitem{Nayak1999} C. Nayak, M.P.A. Fisher, A.W.W. Ludwig, and H.H.
Lin, Phys. Rev. B \textbf{59}, 15694 (1999). 

\bibitem{Chen2002} S. Chen, B. Trauzettel, and R. Egger, Phys. Rev.
Lett. \textbf{89}, 226404 (2002). 

\bibitem{Chamon2003} C. Chamon, M. Oshikawa, and I. Affleck, Phys.
Rev. Lett. \textbf{91}, 206403 (2003). 

\bibitem{Barnabe2005} X. Barnab{é}-Th{é}riault, A. Sedeki, V.
Meden, and K. Sch{ö}nhammer, Phys. Rev. Lett. \textbf{94}, 136405
(2005). 

\bibitem{Oshikawa2006} M. Oshikawa, C. Chamon, and I. Affleck, J.
Stat. Mech.: Theor. Exp. \textbf{P02008} (2006). 

\bibitem{Das2006} S. Das, S. Rao, and D. Sen, Phys. Rev. B \textbf{74},
045322 (2006). 

\bibitem{Hou2008} C.-Y. Hou and C. Chamon, Phys. Rev. B \textbf{77},
155422 (2008). 

\bibitem{Agarwal2009} A. Agarwal, S. Das, S. Rao, and D. Sen, Phys.
Rev. Lett. \textbf{103}, 026401 (2009). 

\bibitem{Giuliano2009} D. Giuliano and P. Sodano, Nucl. Phys. \textbf{B811},
395 (2009). 

\bibitem{Bellazzini2009} B. Bellazzini, P. Calabrese, and M. Mintchev,
Phys. Rev. B \textbf{79}, 085122 (2009). 

\bibitem{Altland2012} A. Altland, Y. Gefen, and B. Rosenow, Phys.
Rev. Lett. \textbf{108}, 136401 (2012). 

\bibitem{Rahmani2012} A. Rahmani, C.-Y. Hou, A. Feiguin, M. Oshikawa,
C. Chamon, and I. Affleck, Phys. Rev. B \textbf{85}, 045120 (2012). 

\bibitem{Altland2015} A. Altland, Y. Gefen, and B. Rosenow, Phys.
Rev. B \textbf{92}, 085124 (2015). 

\bibitem{Yi1998} H. Yi and C.L. Kane, Phys. Rev. B \textbf{57}, R5579(R)
(1998). 

\bibitem{Yi2002} H. Yi, Phys. Rev. B 65, 195101 (2002). 

\bibitem{Ganeshan2016} S. Ganeshan and M. Levin, Phys. Rev. B \textbf{93},
075118 (2016). 

\bibitem{Schrieffer1966} J.R. Schrieffer and P.A. Wolff, Phys. Rev.
\textbf{149}, 491 (1966). 

\bibitem{Cardy1996} J. Cardy, \textit{Scaling and Renormalization
in Statistical Physics} (Cambridge University Press, UK, 1996). 

\bibitem{Protopopov2017} I.V. Protopopov, Y. Gefen, and A.D. Mirlin,
Annals of Physics \textbf{385}, 287 (2017). 

\bibitem{Francesco1997} P. Francesco, P. Mathieu, and D. S{é}n{é}chal,
\textit{Conformal Field Theory} (Springer, New York, 1997). 
\end{thebibliography}
\end{document}